\documentclass[preprintnumbers,prd,twocolumn,showpacs,floatfix,
nofootinbib,superscriptaddress]{revtex4}
\usepackage{graphicx}
\usepackage{epsfig}
\usepackage{bm}
\usepackage{amssymb}
\usepackage{float}
\usepackage{amsmath}
\usepackage{dcolumn}
\usepackage{cancel}
\usepackage[colorlinks]{hyperref}
\usepackage[usenames,dvipsnames]{color}
\hypersetup{
     breaklinks=true,
    pdfstartview={FitH},    
    colorlinks=true,       
    linkcolor=blue,          
    citecolor=red,        
    filecolor=magenta,      
    urlcolor=blue,           
    anchorcolor=green,      
    linktocpage=true
}

\newcommand{\Mpl}{M_{\textrm{Pl}}}
\renewcommand{\(}{\left(}
\renewcommand{\)}{\right)}

\newcommand{\nn}{\nonumber}
\def\al{\alpha}
\def\bet{\beta}
\def\gam{\gamma}

\def\Om{\Omega}
\def\sig{\sigma}
\def\lam{\lambda}
\def\ep{\epsilon}

\def\lap{\nabla}
\def\S{\mathcal{S}}
\def\A{\mathcal{A}}
\def\B{\mathcal{B}}
\def\C{\mathcal{C}}
\def\N{\mathcal{N}}

\def\doi{http://doi.org}
\def\arxiv{http://arxiv.org/abs}
 \def\t{\tilde}
\newcommand{\dc}{\dot{\chi}}

\begin{document}

\title{Variable gravity: A suitable framework for quintessential
inflation
}

\author{Md. Wali Hossain}
 \email{wali@ctp-jamia.res.in}
\affiliation{Centre for Theoretical Physics, Jamia Millia Islamia, New
Delhi-110025,
India}

\author{R. Myrzakulov}\email{rmyrzakulov@gmail.com}
\affiliation{ Eurasian  International Center for Theoretical
Physics, Eurasian National University, Astana 010008, Kazakhstan}

\author{M.~Sami}
\email{sami@iucaa.ernet.in}
\affiliation{Centre for Theoretical Physics,
Jamia Millia Islamia, New Delhi-110025, India}

\author{Emmanuel N. Saridakis}
\email{Emmanuel\_Saridakis@baylor.edu}
\affiliation{Physics Division, National Technical University of Athens,
15780 Zografou Campus,  Athens, Greece}
\affiliation{Instituto de F\'{\i}sica, Pontificia Universidad de Cat\'olica
de Valpara\'{\i}so, Casilla 4950, Valpara\'{\i}so, Chile}

\begin{abstract}

 In this paper, we investigate a scenario of variable gravity and
 apply
it to the unified description of inflation and late time cosmic
acceleration dubbed quintessential inflation. The scalar field
called ``cosmon" which in this model unifies both the concepts
reduces to inflaton at early epochs. We calculate the slow-roll
parameters, the Hubble parameter at the end of inflation, the
reheating temperature,the tensor-to-scalar ration and demonstrate
the agreement of the model with observations and the Planck data.
 As for the post inflationary dynamics, cosmon tracks
the background before it exits the scaling regime at late times. The
scenario gives rise to correct epoch sequence of standard cosmology,
namely, radiative regime, matter phase and dark-energy. We show that
the long kinetic regime after inflation gives rise to enhancement of
relic gravity wave amplitude resulting into violation of
nucleosynthesis constraint at the commencement of radiative regime
in case of an inefficient reheating mechanism such as gravitational
particle production. Instant preheating is implemented to
successfully circumvent the problem. As a generic feature, the
scenario gives rise to a blue spectrum for gravity waves on scales
smaller than the comoving horizon scale at the commencement of the
radiative regime.

\end{abstract}


\maketitle

\section{Introduction}

Theoretical and observational consistencies demand that the
standard model of Universe be complemented by
early phase of accelerated expansion, dubbed inflation 
\cite{Starobinsky:1980te,Starobinsky:1982ee,Guth:1980zm,Linde:1983gd,Linde:1981mu,Liddle:1999mq,Langlois:2004de,Lyth:1998xn,Guth:2000ka,Lidsey:1995np,Bassett:2005xm,Mazumdar:2010sa,Wang:2013hva,Mazumdar:2013gya}, and late time
cosmic acceleration \cite{Copeland:2006wr,Sahni:1999gb,
Frieman:2008sn,Padmanabhan:2002ji,Padmanabhan:2006ag,
Sahni:2006pa,Peebles:2002gy,Perivolaropoulos:2006ce,
Sami:2009dk,Sami:2009jx,Sami:2013ssa}. 
Inflation is a remarkable paradigm, a single
simple idea which addresses logical consistencies of hot Big Bang
and provides   a mechanism for primordial perturbations needed
to seed the structures in the Universe. As for late time cosmic acceleration,
it is now accepted as an observed phenomenon though its underlying
cause still remains to be obscure, whereas similar confirmation of
inflation is still awaited. Thus, the hot Big Bang and the
two phases of accelerated expansion is a theoretically accepted
framework for the description of our Universe.

 No doubt that inflation is a great idea, the phenomenon should therefore
live for ever such that the late time cosmic acceleration is nothing
but its reincarnation {\it \`a la} quintessential inflation
\cite{Peebles:1999fz,Sahni:2001qp,Sami:2004xk,Copeland:2000hn,Huey:2001ae,
Majumdar:2001mm,Dimopoulos:2000md,Sami:2003my,
Dimopoulos:2002hm,Dias:2010rg,BasteroGil:2009eb,Chun:2009yu,Bento:2008yx,
Matsuda:2007ax,da Silva:2007vt,Neupane:2007mu,Dimopoulos:2007bp,
Gardner:2007ib,Rosenfeld:2006hs,Bueno
Sanchez:2006ah,Membiela:2006rj, Bueno
Sanchez:2006eq,Cardenas:2006py,Zhai:2005ub,Rosenfeld:2005mt,
Giovannini:2003jw,Dimopoulos:2002ug,Nunes:2002wz,Dimopoulos:2001qu,
Dimopoulos:2001ix,Yahiro:2001uh,Kaganovich:2000fc,Peloso:1999dm,
Baccigalupi:1998mn,Hinterbichler:2013we}. The idea was first proposed by Peebles and
Vilenkin in 1999 \cite{Peebles:1999fz} which was later implemented
in the framework of braneworld cosmology
\cite{Sahni:2001qp,Sami:2004xk,Copeland:2000hn,Huey:2001ae,
Majumdar:2001mm}. At the theoretical level, it
sounds pretty simple to implement such a proposal in the language of
a single scalar field. The field potential should be shallow at
early times, facilitating slow roll, followed by steep behavior
thereafter and turning  shallow again at late times. The steep
potential is needed for radiative regime to commence, such that the
field is sub-dominant during radiation era and does not interfere
with nucleosynthesis. It should continue to remain in hiding during
matter phase, till its late phases, in order not to obstruct
structure formation. It is then desirable to have scaling regime, in
which the field mimics the background being invisible, allowing the
dynamics to be free from initial conditions, which in turn require a
particular steep behavior of the potential. At late times the field
should overtake the background, giving rise to late-time cosmic
acceleration, which is the case if slow roll is ensured or if the
potential mimics shallow behavior effectively.

There are several obstacles in implementing the above unification
scheme. First, since  inflation survives in this scenario until late
times, the potential is typically of a run-away type and one
therefore requires an alternative mechanism of reheating in this
case. One could invoke reheating due to gravitational particle
production after inflation
\cite{Kofman:1994rk,Dolgov:1982th,Abbott:1982hn,
Ford:1986sy,Spokoiny:1993kt,Kofman:1997yn,Shtanov:1994ce,Campos:2002yk},
which is a universal phenomenon. However, the latter is an inefficient
process and it might take very long for radiative regime to
commence. Clearly, in this case, the scalar field spends long time
in the kinetic regime such that the field energy density redshifts
with the scale factor as $a^{-6}$ corresponding to equation of state
of stiff matter. It is known that the amplitude of gravitational
waves produced at the end of inflation enhances during kinetic
regime, and if the latter is long, the relic gravitational waves
\cite{Sahni:2001qp,Sami:2004xk,Grishchuk:1974ny,Grishchuk:1977zz,
Starobinsky:1979ty,Sahni:1990tx,Souradeep:1992sm,
Giovannini:1998bp,Giovannini:1999bh,Langlois:2000ns,Kobayashi:2003cn,
Hiramatsu:2003iz,Easther:2003re,Brustein:1995ah,Gasperini:1992dp,
Giovannini:1999qj,Giovannini:1997km,Gasperini:1992pa,Giovannini:2009kg,
Giovannini:2008zg,Giovannini:2010yy,Tashiro:2003qp} might come into
conflict with nucleosynthesis constraint at the commencement of
radiative regime \cite{Sahni:2001qp,Sami:2004xk,Tashiro:2003qp}.
Hence, one should look yet for another alternative reheating
mechanism, such as instant preheating
\cite{Felder:1998vq,Felder:1999pv,Campos:2004nc}, to circumvent the
said problem.

A second obstacle to the unification is that if we want the scalar
field to mimic the background for most of the thermal history, the
field potential should behave like a steep exponential potential at
least approximately such as the inverse power-law potentials. Since
scaling regime is an attractor in such cases, an exit mechanism from
scaling regime to late time acceleration should be in place in the
scenario.

Let us examine as how to build the unified picture. The single
scalar field models aiming for quintessential inflation can be
broadly put into two classes: (1) Models in which the field
potential has a required steep behavior for most of the history of
universe but turn shallow at late times, for instance, the inverse
power-law potentials
\cite{Sahni:2001qp,Sami:2004xk,Copeland:2000hn,Sami:2003my}. (2)
Models in which the field potential is shallow at early epochs
giving rise to inflation, followed by the required steep behavior.

In the first class of potentials, we can not implement inflation in
the standard framework, since slow roll needs to be assisted in this
case. For example, one could invoke Randall-Sundrum (RS) braneworld
\cite{Randall:1999ee,Randall:1999vf} corrections
\cite{Sahni:2001qp,Sami:2004xk,Copeland:2000hn} to facilitate
inflation with steep potential at early epochs. In this case, as the
field rolls down to low energy regime, the braneworld corrections
disappear, giving rise to a graceful exit from inflation and
thereafter the scalar field has the required behavior. However,
gravitational particle production \cite{Ford:1986sy,Spokoiny:1993kt}
is extremely inefficient in the braneworld inflation
\cite{Sahni:2001qp} and one could in principle introduce the instant
preheating to tackle the relic gravitational waves problem
\cite{Sami:2004xk}. Unfortunately, the steep braneworld inflation is
inconsistent with observations, namely, the tensor-to-scalar ratio
of perturbations is too high in this case.
Thus, the scenario fails in the early phase, although the late-time
evolution is compatible with theoretical consistency and
observational requirements \cite{Sahni:2001qp,Sami:2004xk}.

In the second class of potentials, that is shallow at early epochs
followed by steep behavior, we need a mechanism to exit from scaling
regime. A possible way out is provided by introducing neutrino
matter, such that neutrino masses are field-dependent
\cite{Wetterich:2013aca,Wetterich:2013jsa,Wetterich:2013wza}. Such a
scenario can be motivated from Brans-Dicke framework, with an
additional assumption on the matter Lagrangian in the Jordan frame,
namely treating massive neutrinos differently from other forms of
matter in a way that the field is minimally coupled to cold dark
matter/baryon matter in Einstein frame whereas the neutrino masses
grow with the field \cite{Wetterich:2013jsa}. In such a scenario
neutrinos do not show up in radiation era; their energy density
tracks radiation being sub-dominant. However, in the subsequent
matter phase at late times, as they become non-relativistic, their
masses begin to grow and their direct coupling to scalar field
builds up such that the effective potential acquires a minimum at
late times giving rise to late time acceleration, provided the field
rolls slowly around the effective minimum. At this point,
 a question arises, namely whether we could do
without neutrino matter and the extra assumption in which case the
field would couple to matter directly in the Einstein frame and the
effective potential would also acquire a minimum. For simplicity let
us assume that we are dealing with a constant coupling $Q$ {\it \`a
la} coupled quintessence \cite{Amendola:1999er}.
In that case it is possible to achieve
slow roll around the minimum of the effective potential, provided
that $Q$ is much larger than the slope of the potential, such that
the effective equation-of-state parameter has a desired negative
value ($w_{\rm eff}=(-Q+\al)/(Q+\al)$ where $\al$ is the slope of
the potential). The scaling solution (which is accelerating thanks to
non-minimal coupling), an attractor of the dynamics, is approached
soon after the Universe enters into matter-dominated regime and
consequently we cannot have a viable matter phase in this case. It
is therefore necessary that the matter regime is left intact and the
transition to accelerated expansion takes place only at late times.
The latter can be triggered by massive neutrino matter with
field-dependent masses
\cite{Amendola:2007yx,Wetterich:2007kr,Pettorino:2010bv}.

In this paper we consider a scenario of quintessential inflation in
the framework of variable gravity model
\cite{Wetterich:2013jsa,Wetterich:2013aca,Wetterich:2013wza,
Wetterich:2014eaa,Wetterich:2014bma}. 
We first revisit the
model in Jordan frame (Sec.~\ref{JF}) and then we transit to the
Einstein frame (Sec.~\ref{EF}) for
detailed investigations of cosmological dynamics by considering the
canonical form of the action (Sub Sec.~\ref{Canonical}). Behavior
of the canonical field with respect to the non canonical field
is also examined (Sub Sec.~\ref{sig_behavior}). In the Einstein
frame we examine the inflationary phase (Sec.~\ref{Inflation}),
kinetic regime and late
time transition to dark energy (Sec.~\ref{Late times evolution}).
Ref.\cite{Wetterich:2013jsa} provides broad out line of inflation and late time acceleration in the framework of model under consideration. In this paper, we present complete evolution history by invoking suitable preheating mechanism.
We investigate issues related to the
spectrum of relic gravity waves (Sub Sec.~\ref{RGW}) as a generic observational features
of quintessential inflation. The relic gravity wave amplitude is
defined by the inflationary Hubble parameter whereas the spectrum of
the wave crucially depends upon the post inflationary evolution. We
investigate the problems related to the long kinetic regime in the
scenario and discuss the instant preheating
(Sub Sec.~\ref{Instant Preheating}) to tackle the problem. Post
inflationary evolution (Sub Sec.~\ref{Cosmological equations}) is
investigated with canonical action and the epoch sequences
(Sub Sec.~\ref{Transient Behavior}) are achieved with viable matter
phase. Detailed dynamical analysis is performed to check the nature of
stability of all fixed points (Sub Sec.~\ref{Asymptotic behavior}). Finally
in Sec.~\ref{Conclusions} we summarize the results.

\section{Variable Gravity in Jordan Frame}
\label{JF}
In this section we revisit and analyze the variable gravity
model\cite{Wetterich:2013jsa,Wetterich:2013aca} to be used for our
investigations. The scenario of variable gravity is characterized by
the following action in the Jordan-frame, :
 \begin{eqnarray}
\label{eq:action_J}
\mathcal{S}_J &=& \int  d^4 x
\sqrt{\tilde g}\left[-\frac{1}{2}\tilde F(\chi)\tilde R+\frac{1}{2}
\tilde K(\chi)\partial^\mu\chi\partial_\mu\chi+\tilde V(\chi)\right] \nn \\ &&+
\t \S_m+\t\S_r+\t\S_\nu \, ,
\end{eqnarray}
where tildes represent the quantities in the Jordan frame. In the
above action $\chi$ is the cosmon field with $\tilde V(\chi)$, and
apart from the coupling $\tilde K(\chi)$ we have considered an
effective Planck mass $\tilde F(\chi)$ driven by the field.
Additionally, $\mathcal{\tilde S}_m$ and $\mathcal{\tilde S}_r$ are
the matter and radiation actions respectively and $\tilde\S_\nu$ is
the action for neutrino matter, which we have considered separately
since massive neutrinos play an important role in this model during
late times. During the radiation era or earlier, neutrinos are ultra
relativistic or relativistic, which implies that neutrinos behave as
radiation during and before radiation era, with their mass being
constant. After the radiation era neutrinos start losing their
energy and become non-relativistic, behaving like ordinary matter
with zero pressure. During late times neutrino mass starts growing
with the field, and along with the cosmon field $\chi$ they give rise to
late-time de Sitter solution \cite{Fardon:2003eh,Bi:2003yr,
Hung:2003jb,Peccei:2004sz,Bi:2004ns,Brookfield:2005td,Brookfield:2005bz,
Amendola:2007yx,Bjaelde:2007ki,Afshordi:2005ym,Wetterich:2007kr,Mota:2008nj,
Pettorino:2010bv,LaVacca:2012ir,Collodel:2012bp}.

In this construction the variation in the particles masses comes
from the non-minimal coupling of the field with matter. For
radiation this non-minimal coupling does not affect its continuity
equation, since the energy-momentum tensor for radiation is
traceless. We consider different couplings of the field with matter,
radiation, and neutrinos, that is we consider the non-minimal
coupling $\A^2(\chi)$ between the cosmon field and matter, and the
non-minimal coupling $\B^2(\chi)$ between the cosmon field and the
neutrinos. Without loss of generality we consider the non-minimal
coupling between the field and radiation to be $\A(\chi)^2$ too. To
sum up, we shall use the following actions,
\begin{eqnarray}
 \tilde\S_m &=& \tilde\S_m(\A^2\tilde g_{\alpha\beta};\Psi_m) \, , \\
 \tilde\S_r &=& \tilde\S_r(\A^2\tilde g_{\alpha\beta};\Psi_r) \, , \\
 \tilde\S_\nu &=& \tilde\S_\nu(\B^2\tilde g_{\alpha\beta};\Psi_\nu) \, .
\end{eqnarray}

Variation of the action (\ref{eq:action_J}) with respect to the metric leads
to the Einstein field equation
\begin{eqnarray}
 \t F\(\t R_{\al\bet}-\frac{1}{2}\t R\t g_{\al\bet}\) &=&\t K\partial_\al\chi\partial_\bet\chi
-\frac{1}{2}\t K\t g_{\al\bet}\partial^\rho\chi\partial_\rho\chi
 \nn \\ &&
 -\t V\t g_{\al\bet}+\t\lap_\al\t\lap_\bet \t F-\t\lap^2 \t F g_{\al\bet} \nn \\ 
 && +\t T_{\al\bet}  = \t F \t G_{\al\bet} \, ,
 \label{eq:Eins_J}
\end{eqnarray}
where $\t T_{\al\bet}$ includes the contributions from matter, radiation and
neutrinos, that is
$\t T_{\al\beta}=\t T^{(m)}_{\al\beta}+\t T^{(r)}_{\al\beta}+\t
T^{(\nu)}_{\al\beta}$.

Variation of action (\ref{eq:action_J}) with respect to the cosmon field $\chi$
provides its equation of motion of the field, namely
\begin{equation}
 \t K\t\Box \chi+\frac{1}{2}\frac{\partial \t K}{\partial\chi}\partial^\mu\chi\partial_\mu\chi=
 \frac{\partial \t V}{\partial \chi}-\frac{1}{2}\frac{\partial \t F}{\partial\chi}\t R+\t q_\chi \, ,
 \label{eq:eom_chi}
\end{equation}
where $\t q_\chi=\t q_{\chi,m}+\t q_{\chi,\nu}+\t q_{\chi,r}$ and
\begin{eqnarray}
 \t q_{\chi,m} &=& \frac{1}{\sqrt{-\tilde g}} \frac{\delta\tilde\S_m}{\delta\chi}
              = \frac{\A'}{\A}\t T^{(m)}= - \frac{\partial\ln \A}{\partial \chi}(\t\rho_m-3\t p_m)
                , \,\,\,\,\,\,
  \label{eq:q_chim}\\
 \t q_{\chi,\nu} &=& \frac{1}{\sqrt{-\tilde g}} \frac{\delta\tilde\S_\nu}{\delta\chi}
               =\frac{\B'}{\B}\tilde T^{(\nu)} =- \frac{\partial\ln \B}{\partial \chi}\(\tilde\rho_\nu-3\tilde p_\nu\)
                \, ,
 \label{eq:q_chinu}\\
 \t q_{\chi,r} &=& \frac{1}{\sqrt{-\tilde g}} \frac{\delta\tilde\S_r}{\delta\chi} = 0 \, .
\end{eqnarray}
Here the primes represent derivatives with respect to $\chi$ and the
energy-momentum tensors are defined as
\begin{equation}
 \tilde T_{\alpha\beta}^{(i)}=-\frac{2}{\sqrt{-\tilde g}} \frac{\delta\tilde\S_i}{\delta\tilde g^{\alpha\beta}} \, .
\end{equation}
One can easily see that \cite{Wetterich:2013jsa}
\begin{eqnarray}
 \t q_{\chi,m} &=& -\frac{\partial \ln m_p}{\partial\chi}\(\tilde\rho_m-3\tilde p_m\) = -\frac{m_pn_p}{\chi}\, ,
 \label{eq:q_chim1} \\
 \t q_{\chi,\nu} &=& -\frac{\partial \ln m_\nu}{\partial\chi}\(\tilde\rho_\nu-3\tilde p_\nu\)
 = -\(2\tilde{\gamma}+1\)\frac{m_\nu n_\nu}{\chi} \, ,\,\,\,\,\,\,
 \label{eq:q_chinu1}
\end{eqnarray}
where $m_p$ and $m_\nu$ are the masses of matter-particles and neutrino and
$n_p$ and $n_\nu$ are the number densities of the matter-particles and
the neutrinos respectively. In the above expression we have followed
\cite{Wetterich:2013jsa}  and for convenience we have
defined  $\tilde{\gamma}$ through
\begin{eqnarray}
\label{gamatilde}
m_\nu\propto\chi^{2\tilde{\gamma}+1}
\end{eqnarray}

Comparing  (\ref{eq:q_chim1}) and  (\ref{eq:q_chinu1}) with
(\ref{eq:q_chim})
and  (\ref{eq:q_chinu}) respectively, we can see that $m_p\sim \A$ and
$m_\nu\sim\B$. Thus, choosing suitable $\A$ and $\B$ we can match our
considerations with those of \cite{Wetterich:2013jsa}. In particular,
according to \cite{Wetterich:2013jsa} particles masses vary linearly with
the cosmon field, apart from the neutrinos. That is
$\A(\chi)^2=\chi^2/\Mpl^2$, which leads to $m_p\sim\A\sim\chi$. Neutrino
mass varies slightly differently than the other particles.
In particular, $\B(\chi)^2=\left(\chi/\Mpl\right)^{4\tilde{\gamma}+2}$,
which gives   $m_\nu\sim\B\sim\chi^{2\tilde{\gamma}+1}$, with
$\tilde{\gamma}$  a
constant.

We shall consider four matter components in the universe,namely,
radiation, baryonic+cold dark matter (CDM), neutrinos and the
contribution of the cosmon field. Furthermore, we stress that the
late-time dark energy is attributed to two contributions, namely to
both the cosmon and the neutrino fields. Thus, the total
energy-momentum tensor, which can be calculated from action
(\ref{eq:action_J}), reads
\begin{eqnarray}
 \tilde T_{\alpha\beta}=\tilde T^{(m)}_{\alpha\beta}+\tilde T^{(r)}_{\alpha\beta}+
 \tilde T^{(\nu)}_{\alpha\beta}+\tilde T^{(\chi)}_{\alpha\beta} \, ,
\end{eqnarray}
where
\begin{eqnarray}
 \tilde T_{\al\beta}^{(\chi)} &=& \t K\partial_\al\chi\partial_\beta\chi-
 \tilde g_{\al\beta}\(\frac{1}{2}\t K\partial^\rho\chi\partial_\rho\chi+\t V\)\nn \\ &&
 +\t\lap_\al\t\lap_\beta F-\t\lap^2 \t F\t g_{\al\beta}+\(\t F_0-\t F\)\t G_{\al\beta} \, ,
 \label{eq:emchi}
\end{eqnarray}
with $\t F_0=\t F(\chi_0)$   the present value of $\t F(\chi)$ \footnote{Eq.
(\ref{eq:emchi}) is calculated
by writing Eq.~(\ref{eq:Eins_J}) as the standard one, that is
\begin{eqnarray}
 \t F_0 \t G_{\al\bet} &=&\t K\partial_\al\chi\partial_\bet\chi
-\frac{1}{2}\t K \t g_{\al\bet}\partial^\rho\chi\partial_\rho\chi
 -\t V\t g_{\al\bet}+\t\lap_\al\t\lap_\bet \t F \nn \\ && -\t\lap^2 \t F \nn-\t F\t G_{\al\bet}+\t T_{\al\bet}  \, ,
\end{eqnarray}
where $F_0$ gives the present value of the Newton's constant.
}.

The evolution equations of the various sectors in the model at hand
read:
 \begin{eqnarray}
 \label{eq:matcont}
\dot{\tilde\rho}_m+3\tilde H(\tilde\rho_m+\tilde p_m)&=&-\t q_{\chi
m}\dot{\chi}=(\tilde\rho_m-3\tilde p_m)\frac{\dc}{\chi} \, , \, \, \\
 \label{eq:neutcont}
\dot{\tilde\rho}_\nu+3\tilde H(\tilde\rho_\nu+\tilde p_\nu)&=&-\t q_{\chi
m}\dot{\chi} \nn \\ &=&\(2\tilde{\gamma}+1\)(\tilde\rho_\nu-3\tilde
p_\nu)\frac{\dc}{\chi} \, , \\
 \label{eq:radcont}
\dot{\tilde\rho}_r+3\tilde H(\tilde\rho_r+\tilde p_r)&=&0 \, ,
\end{eqnarray}
which follow from the equations
\begin{eqnarray}
 \tilde T^{(m)\al}_{~~~~\beta;\al} &=& \t q_{\chi,m} \chi_{,\beta} \, , \\
 \tilde T^{(\nu)\al}_{~~~\beta;\al} &=& \t q_{\chi,\nu} \chi_{,\beta} \, , \\
 \tilde T^{(r)\al}_{~~~\beta;\al} &=& \t q_{\chi,r} \chi_{,\beta} \,.
\end{eqnarray}

From Eqs.~(\ref{eq:matcont}), (\ref{eq:neutcont}) and
(\ref{eq:radcont})  we can extract the
continuity equation for the cosmon field, which writes as
\begin{eqnarray}
   \label{eq:DEcont}
\dot{\t\rho}_\chi+3\t H(\t\rho_\chi+\t p_\chi)&=&\t q_{\chi m}\dot{\chi}
=-\left\{\(\t\rho_m-3\tilde p_m\) \nn \right. \\ && \left. +
(2\t\gam+1)\(\t\rho_\nu-3\t p_\nu\)\right\}\frac{\dc}{\chi}\, . ~~
\end{eqnarray}

Finally, the consistency check of the Eqs.~(\ref{eq:matcont}),
(\ref{eq:neutcont}), (\ref{eq:radcont}) and
(\ref{eq:DEcont}) follows from the conservation equation of the total energy
$\tilde\rho_T=\tilde\rho_m+\tilde\rho_\nu+\tilde\rho_r+\tilde\rho_\chi$:
\begin{eqnarray}
 \dot{\tilde\rho}_T+3\tilde H\(\tilde\rho_T+\tilde p_T\)=0 \, .
\end{eqnarray}

We close this section by mentioning that, although the above model
looks similar to extended quintessence
\cite{Chiba:1999wt,Uzan:1999ch,Baccigalupi:2000je,Amendola:1999er},
or as a special case of the generalized Galileon models , there is a
crucial difference, namely that the particle masses depend on
$\chi$, that is the matter energy density and pressure depend on
$\chi$ too. This has an important phenomenological consequence the
appearance of an effective interaction between the scalar field and
matter and neutrinos, described by relations (\ref{eq:matcont}),
(\ref{eq:neutcont}) and (\ref{eq:DEcont}). In the discussion to
follow, it would be convenient to work in the Einstein frame.

\section{Variable Gravity in Einstein Frame}
\label{EF}

In this section we examine the variable gravity model in the
Einstein frame and analyze the aspects related to early
phase,thermal history and late time evolution. Let us consider the
following conformal transformation,
\begin{equation}
 g_{\mu\nu}=\Omega^2 \tilde g_{\mu\nu} \, ,
 \label{eq:ct}
\end{equation}
where $\Omega^2=\tilde F(\chi)/\Mpl^2$ is the conformal factor and
$g_{\mu\nu}$ is the Einstein-frame metric.

Using the conformal transformation (\ref{eq:ct}) one can easily show that
\begin{eqnarray}
 &&\t R =\Omega^2\(R+6\Box\ln\Omega-6g^{\mu\nu}
\partial_\mu\ln\Omega\partial_\nu\ln\Omega\) \nn \\
          &&= \frac{\tilde F}{\Mpl^2}\left\{R+3\Box
          \ln\(\frac{\tilde F}{\Mpl^2}\) -
          \frac{3}{2\tilde F^2}g^{\mu\nu} \nn \right. \\ && \left. \times \partial_\mu\tilde
F\partial_\nu\tilde F\right\} \, , \\ 
  &&\sqrt{-\tilde g} = \Omega^{-4}\sqrt{g}\, .
  \end{eqnarray}
Therefore, under the conformal transformation (\ref{eq:ct}) the Jordan-frame
action (\ref{eq:action_J}) becomes
\begin{eqnarray}
 \label{eq:action_E1}
\mathcal{S}_E &=& \int  d^4 x
\sqrt{g}\Bigg[\Mpl^2\(-\frac{1}{2}R+\frac{1}{2\chi^2}
K(\chi)\partial^\mu\chi\partial_\mu\chi\) \nn \\ && +V(\chi)\Bigg] +
\mathcal{S}_m+\mathcal{S}_r+\S_\nu \, ,
\end{eqnarray}
where,
\begin{eqnarray}
 V(\chi) &=& \frac{\Mpl^4\tilde V}{\tilde F^2} \, , \\
 K(\chi) &=& \chi^2\Big[\frac{\tilde K}{\tilde F}+
 \frac{3}{2}\(\frac{\partial \ln \tilde F}{\partial \chi}\)^2\Big] \, .
\end{eqnarray}

In this work following \cite{Wetterich:2013jsa}, we consider the
choice,
\begin{eqnarray}
&& \t
F(\chi)=\chi^2 \, ,\\
&& \t
K(\chi)=\frac{4}{\t\al^2}\frac{m^2}{\chi^2+m^2}+\frac{4}{\al^2}\frac{\chi^2}{
\chi^2+m^2}-6 \, ,
 \label{eq:K}
\end{eqnarray}
where $\t\al$ and $\al$ are constants (the tilde in $\t\al$  has
nothing to do with the frame choice). The parameter $m$ is an
intrinsic mass scale which plays a crucial role in inflation, when
$\chi\lesssim m$, but can be neglected during and after radiation
era when $\chi$ grows to a higher value such that $\chi\gg m$.
Hence, for the late time behavior of the model we can
  use the
approximation $\chi\gg m$, which gives approximately a constant $\t K(\chi)$:
\begin{eqnarray}
 \t K\approx \frac{4}{\al^2}-6 \, .
 \label{eq:K_approx}
\end{eqnarray}

One can easily see that in the Einstein frame,  neutrino matter is
non-minimally coupled to cosmon field whereas matter and radiation
are minimally coupled. Indeed, we have
\begin{eqnarray}
 \label{eq:sm}
 \S_m &=& \tilde\S_m(\Omega^{-2}\A^2\tilde g_{\alpha\beta};\Psi_m)=
\tilde\S_m(g_{\al\beta};\Psi_m) \, , \\
 \S_r &=& \tilde\S_r(\Omega^{-2}\A^2\tilde g_{\alpha\beta};\Psi_r)=
\tilde\S_r(g_{\al\beta};\Psi_r) \, ,
 \label{eq:sr}\\
 \S_\nu &=& \tilde\S_\nu(\Omega^{-2}\B^2\tilde g_{\alpha\beta};\Psi_\nu)
 = \tilde\S_\nu(\(\chi/\Mpl\)^{4\tilde{\gamma}}g_{\alpha\beta};\Psi_\nu)  .\,
~~~~~
 \label{eq:snu}
\end{eqnarray}
Thus,  from  (\ref{eq:sm}), (\ref{eq:sr}) and (\ref{eq:snu}) we
deduce that only the neutrino mass is field-dependent in the
Einstein frame, while the other particles masses remain constant as
it should be \cite{Wetterich:2013jsa,Wetterich:2013aca}, that is
 \begin{eqnarray}
 \label{eq:matcontef}
&&\dot{\rho}_m+3H(\rho_m+p_m)=0\, ,\\
 \label{eq:radcontef}
&& \dot{\rho}_r+3H(\rho_r+p_r)=0 \, , \\
 \label{eq:neutcontef}
&&\dot{\rho}_\nu+3H(\rho_\nu+p_\nu)=2\tilde{\gamma}(\rho_\nu-3p_\nu)\frac{\dc
}{\chi}
\, .
\end{eqnarray}

Eqs.~(\ref{eq:matcontef}) and (\ref{eq:radcontef}) imply that $\rho_m\sim
a^{-3}$ and $\rho_r\sim a^{-4}$, as usual. However, interestingly enough the
neutrino behavior changes from era to era. During radiation epoch or earlier,
neutrinos behave as radiation, that is the r.h.s. of Eq.
(\ref{eq:neutcontef}) becomes zero and thus $\rho_\nu\sim a^{-4}$. On the
other hand, after the radiation epoch neutrinos start becoming
non-relativistic and behaving like non-relativistic matter, that is
$p_\nu\sim 0$ during and after matter era. However, note that
Eq.~(\ref{eq:neutcontef}) implies that the neutrino mass depends on the field
when the r.h.s. of Eq.~(\ref{eq:neutcontef}) is non zero, and therefore we
deduce that during or after the matter era, the neutrino density $\rho_\nu$
does not evolve as $\sim a^{-3}$.

In order to proceed further, we consider a quadratic potential in
the Einstein frame \cite{Wetterich:2013jsa,Wetterich:2013aca}
\begin{eqnarray}
 \t V(\chi) &=& \mu^2 \chi^2 \, .
\end{eqnarray}
It proves convenient to redefine the field $\chi$ in terms of a new field
$\phi$ as
\begin{equation}
 \chi=\mu ~e^{\frac{\alpha\phi}{2\Mpl}} \,.
\end{equation}
In this case action (\ref{eq:action_E1}) becomes
\begin{eqnarray}
 \label{eq:action_E2}
\S_E &=& \int  d^4 x\sqrt{g}\left[-\frac{\Mpl^2}{2}R+\frac{1}{2}
k^2(\phi)\partial^\mu\phi\partial_\mu\phi+V(\phi)\right] \nn \\ &&+
\S_m+\S_r+\S_\nu(\C^2g_{\al\beta};\Psi_\nu)  , ~~~~~
\end{eqnarray}
with
\begin{eqnarray}
\label{kphi}
 k^2(\phi)&=&\frac{\alpha^2(\tilde
K+6)}{4}=\frac{\alpha^2+\t\alpha^2\mu_m^2e^{\frac{\alpha\phi}{\Mpl}}}{\t\al^2
\left(\mu_m^2e^{\frac{\alpha\phi}{\Mpl}}+1\right)} \\
 \C(\phi)^2&=& (\mu/\Mpl)^{4\tilde{\gamma}}e^{2\tilde{\gamma}\al\phi/\Mpl},
\end{eqnarray}
where  we have defined $\mu_m\equiv\mu/m$, which according
to \cite{Wetterich:2013jsa} $\mu_m\approx 0.01$.
Let us note that the action (\ref{eq:action_E2})is a particular case of Horndeski class with higher derivative terms absent and the coefficient of kinetic term having dependence on the field $\phi$ alone.
 Secondly, the system is free of ghosts as $k(\phi)$ is positive definite in our choice.

Variation of the action (\ref{eq:action_E2}) with respect to the metric gives
\begin{eqnarray}
 \Mpl^2 G_{\al\beta}&=&\Mpl^2\(R_{\al\beta}-\frac{1}{2}Rg_{\al\beta}\) \nn \\ &=& T_{\al\beta}^{(\phi)}+
 T_{\al\beta}^{(m)}+T_{\al\beta}^{(r)} +T_{\al\beta}^{(\nu)}  \, ,
 \label{eq:Eins}
\end{eqnarray}
where
\begin{eqnarray}
 T^{(\phi)}_{\mu\nu}=-\frac{1}{2}k^2g_{\mu\nu}\partial^\rho\phi\partial_\rho\phi+
 k^2\partial_\mu\phi\partial_\nu\phi-V(\phi)g_{\mu\nu} \, ,
\end{eqnarray}
and
\begin{equation}
 V(\phi)=\Mpl^4e^{-\alpha\phi/\Mpl} \, .
 \label{eq:pot}
\end{equation}
Moreover, variation of (\ref{eq:action_E2}) with respect to the
re-scaled cosmon field $\phi$ gives its equation of motion, namely
\begin{equation}
 k^2\Box \phi+k\frac{\partial k}{\partial\phi}\partial^\mu\phi\partial_\mu\phi=
 \frac{\partial V}{\partial
\phi}+\frac{\tilde{\gamma}\al}{\Mpl}\(\rho_\nu-3p_\nu\) \, .
 \label{eq:eom_phi}
\end{equation}
Finally, note that in terms of the field $\phi$ Eq.~(\ref{eq:neutcontef})
becomes,
\begin{eqnarray}
  \label{eq:neutcontefphi}
\dot{\rho}_\nu+3H(\rho_\nu+p_\nu)=\tilde{\gamma}\al(\rho_\nu-3p_\nu)\frac{
\dot\phi}{
\Mpl} \, ,
\end{eqnarray}
which can then be re-expressed in terms of the neutrino mass $m_\nu$ as
\cite{Brookfield:2005td,Brookfield:2005bz}
\begin{eqnarray}
  \label{eq:neutcontefphi1}
\dot{\rho}_\nu+3H(\rho_\nu+p_\nu)=\frac{\partial \ln m_\nu}
{\partial\phi}(\rho_\nu-3p_\nu)\dot\phi \, .
\end{eqnarray}
Thus, comparing Eq.~(\ref{eq:neutcontefphi}) and Eq.
(\ref{eq:neutcontefphi1}) we deduce that
\begin{equation}
 m_\nu=m_{\nu,0}e^{\tilde{\gamma}\al\phi/\Mpl} \, ,
 \label{eq:mnu_phi}
\end{equation}
where $m_{\nu,0}=m_\nu(\phi=0)=m_\nu(\chi=\mu)$. Since at the present time
$\chi\approx\Mpl$ we can write,
\begin{eqnarray}
 m_{\nu,0}=m_\nu(z=0)\times\(\frac{\mu}{\Mpl}\)^{2\t\gam} \, ,
 \label{eq:m_nu_0}
\end{eqnarray}
where $z$ is the redshift and $m_\nu(z=0)$ is the present value of the
neutrino mass.

\begin{figure}[h]
\centering
\includegraphics[scale=.75]{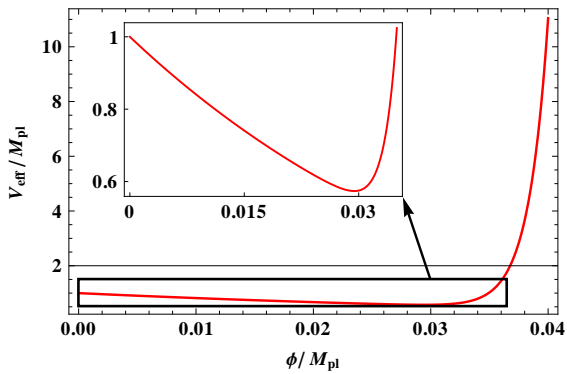}
\caption{Effective potential (\ref{eq:pot_eff_phi}) is plotted against the non-canonical field $\phi$. $\t\gam=30\;, \al=20\;, \hat\rho_\nu/\Mpl^4=10^{-10}\; {\rm and}\; \hat p_\nu/\Mpl^4=0$ are the chosen values of different parameters. Here we should note that if we change the values of the parameters $\hat\rho_\nu/\Mpl^4$ and $\hat p_\nu/\Mpl^4$ then the nature of the effective potential does not change and only the position of the minimum shifts along the horizontal axis. The inset shows the minimum of the effective potential.}
\label{fig:eff_pot}
\end{figure}

Finally, from the r.h.s. of Eq.~(\ref{eq:eom_phi}) we can define the
effective potential
\begin{eqnarray}
 V_{\rm eff}(\phi)=V(\phi)+\(\hat\rho_\nu-3\hat
p_\nu\)e^{(\tilde{\gamma}\al\phi/\Mpl)}  \, ,
 \label{eq:pot_eff_phi}
\end{eqnarray}
where $\hat\rho_\nu=\rho_\nu e^{-(\tilde{\gamma}\al\phi/\Mpl)}$ and
$\hat p_\nu=p_\nu e^{-(\tilde{\gamma}\al\phi/\Mpl)}$ are independent of
$\phi$.
This effective potential $V_{\rm eff}$ has a minimum at
\begin{eqnarray}
 \phi_{\rm min}=\frac{\Mpl}{\al(1+\tilde{\gamma})}\ln\left[\frac{\Mpl^4}
{\tilde{\gamma}(\hat\rho_\nu-3\hat p_\nu)}\right] \, ,
\label{eq:phi_min}
\end{eqnarray}
which is the key feature in the scenario under consideration. By
setting the model parameters, it is possible to achieve minimum at
late times such that the field rolls slowly around the minimum of
the effective potential. The role of neutrino matter is solely
related the transition to stable de Sitter around the present epoch. Fig.~\ref{fig:eff_pot} shows the nature of the effective potential (\ref{eq:pot_eff_phi}) and the inset shows the minimum of the effective potential. 

Using Eq.~(\ref{eq:phi_min}) we get the minimum value of the effective
potential (\ref{eq:pot_eff_phi}) for $\phi=\phi_{\rm min}$,
\begin{eqnarray}
 V_{\rm eff,min}=\(1+\frac{1}{\t\gam}\)V_{\rm min} \, ,
 \label{eq:Veff_min}
\end{eqnarray}
where $V_{\rm min}=V(\phi_{\rm min})$.

Eq.~(\ref{eq:Veff_min}) can be represented in terms of the neutrino
mass by using Eq.~(\ref{eq:mnu_phi}) and Eq.~(\ref{eq:m_nu_0}),
\begin{eqnarray}
 V_{\rm eff,min}=\(1+\frac{1}{\t\gam}\)\(\frac{m_\nu(z=0)}{m_{\nu,\rm min}}\)^{1/\t\gam}\mu^2\Mpl^2 \, ,
 \label{eq:Veff_min_1}
\end{eqnarray}
where $m_{\nu,\rm min}=m_\nu(\phi=\phi_{\rm min})$.

Now $\mu\approx H_0$ \cite{Wetterich:2013jsa} where $H_0$ is the present value of the
Hubble parameter. So in Eq.~(\ref{eq:Veff_min_1}) the term $\mu^2\Mpl^2\sim H_0^2\Mpl^2$.
If we take $H_0=70\rm kmMpc^{-1}sec^{-1}$ then we have in eV, $H_0=1.5\times 10^{-33}\rm eV$.
So $V_{\rm eff,min}$ will be of the order of $3H_0^2\Mpl^2$ only when,
\begin{eqnarray}
 \(1+\frac{1}{\t\gam}\)\(\frac{m_\nu(z=0)}{m_{\nu,\rm min}}\)^{1/\t\gam}\approx 1 \, ,
\end{eqnarray}
and since to get late time cosmic acceleration field has to settle down at the
minimum of the effective potential during the present time we can safely take
$m_\nu(z=0)=m_{\nu,\rm min}$, which implies $\t\gam\gg 1$.

\subsection{Canonical form of the action}
\label{Canonical}

Let us now transform the scalar-field part of the action
(\ref{eq:action_E2}) to its canonical form through the
transformation
\begin{eqnarray}
\label{sigmadef}
 \sigma &=& \Bbbk(\phi) \, ,\\
 k^2(\phi) &=& \(\frac{\partial\Bbbk}{\partial \phi}\)^2 \, ,
 \label{eq:dsig}
\end{eqnarray}
where $k^2(\phi)$ is given by (\ref{kphi}).
Thus,    (\ref{eq:action_E2}) becomes
\begin{eqnarray}
 \label{eq:action_E3}
\mathcal{S}_E &=& \int  d^4 x\sqrt{g}\left[-\frac{\Mpl^2}{2}R+\frac{1}{2}
\partial^\mu\sigma\partial_\mu\sigma+V(\Bbbk^{-1}(\sigma))\right] \nn \\ && +
\mathcal{S}_m+\mathcal{S}_r+\S_\nu(\C^2g_{\al\bet};\Psi_\nu) \, ,
\end{eqnarray}
where $\C(\sig)$ is the conformal coupling in the Einstein frame between the
canonical field $\sig$ and neutrinos. As we can see, the scalar field has
now the canonical kinetic term.

The $\phi$ dependence of the canonical field $\sigma$ can be calculated from
Eq.~(\ref{eq:dsig}), and writes as
\begin{eqnarray}
 \frac{\sigma(\phi)}{\Mpl} &=& \frac{\alpha\phi}{\tilde\alpha \Mpl}-
 \frac{1}{\tilde\alpha}
 \ln\left\{2\alpha^2+e^{\alpha\phi/\Mpl}\mu_m^2\(\alpha^2+\tilde\alpha^2\)
\nn \right. \\ && \left. +
 2\alpha\sqrt{\(1+e^{\alpha\phi/\Mpl}\mu_m^2\)\(\alpha^2+e^{\alpha\phi/\Mpl}
\mu_m^2\tilde\alpha^2\)}\right\}
 \nn \\ &&
 +\frac{1}{\alpha}\ln\left\{\alpha^2+\tilde\alpha\left[\tilde\alpha+2e^{
\alpha\phi/\Mpl }\mu_m^2\tilde\alpha
 \nn \right. \right. \\ && \left.\left.+
2\sqrt{\(1+e^{\alpha\phi/\Mpl}\mu_m^2\)\(\alpha^2+e^{\alpha\phi/\Mpl}
\mu_m^2\tilde\alpha^2\)}\right]\right\}
\nn \\ &&+C  , ~~~~~
 \label{eq:sig}
 \end{eqnarray}
where  $C$ is an integration constant.
We consider $\sigma(\phi=0)=0$ \footnote{The choice of
$\sig(\phi=0)$ also gives $\chi\to 0$ as $\sig\to -\infty$, similar to the
$\phi$ field. Therefore, the value of $C$ we are getting here can also be
obtained from Eq.~(\ref{eq:sig}) by putting
$e^{\al\phi/\Mpl}= 0$ and considering $\sig(\chi\to 0)=\phi(\chi\to 0)$.},
which gives
\begin{eqnarray}
 && C = \frac{1}{\tilde\alpha}
 \ln\left\{2\alpha^2+\mu_m^2\(\alpha^2+\tilde\alpha^2\)  \nn \right. \\ && \left.+
 2\al\sqrt{\(1+\mu_m^2\)\(\alpha^2+\mu_m^2\tilde\alpha^2\)}\right\}
 \nn \\ &&
 -\frac{1}{\alpha}
\ln\left\{\alpha^2+\tilde\alpha\left[\tilde\alpha+2\mu_m^2\tilde\alpha
 \nn \right.\right. \\ && \left.\left.+
 2\sqrt{\(1+\mu_m^2\)\(\alpha^2+\mu_m^2\tilde\alpha^2\)}\right]\right\}
\label{eq:C0}
\end{eqnarray}
Additionally, if we consider $\t\al$ very small  (according to
\cite{Wetterich:2013jsa}
$\t\al\lesssim 0.02$) and $\al$  large  comparing to $\t\al$ and $\mu_m$
\cite{Wetterich:2013jsa}, we can approximate it as
\begin{eqnarray}
 C \approx  \frac{2}{\t\al}\ln\(2\al\)-\frac{2}{\al}\ln\(\al+\t\al\) \, .
 \label{eq:C}
\end{eqnarray}

Finally, note that in order to write the explicit form of
$V(\sigma)=V(\Bbbk^{-1}(\sigma))$ in (\ref{eq:action_E3}), we need to invert
 (\ref{eq:sig}) in order to obtain the explicit form of $\phi(\sigma)$, and
then substitute into $V(\phi)$ in (\ref{eq:pot}). However, (\ref{eq:sig}) is
a transcendental equation and thus it cannot be inverted. Fortunately, in
the following elaboration $V(\sigma)$ will appear only through its derivative
$dV(\sigma)/d\sigma$, which using  (\ref{sigmadef}), (\ref{eq:dsig})
acquires the simple form
\begin{eqnarray}
\frac{d V(\sigma)}{d\sigma}=\frac{1}{k(\phi)}\frac{d V(\phi)}{d\phi}.
 \label{dVsigma}
\end{eqnarray}
In order to check whether the behavior of the field can comply with
requirements spelled out in the aforesaid discussion, it would be
convenient to check for the asymptotic behavior of the potential.

\subsection{Asymptotic behavior}
\label{sig_behavior}

In the previous subsection we extracted the expressions for $\sigma(\phi)$,
$k(\phi)$ and $dV(\sigma)/d\sigma$, where $\sigma$ is the redefined
scalar field, in terms of which the action takes the canonical form. Since
the involved expressions are quite complicated, it would be useful to obtain
their asymptotic approximations. In particular, we are interested in the
two limiting regimes, that is for small $\chi$ ($\chi\ll m$ or equivalently
$\phi\ll -2 \Mpl\ln (\mu_m)/\al$) and large $\chi$ ($\chi\gg m$ or
equivalently
$\phi\gg -2 \Mpl\ln (\mu_m)/\al$)
respectively.

For small $\chi$   from   (\ref{eq:K}),(\ref{kphi}) we have
\begin{equation}
 k^2(\phi)\approx\frac{\alpha^2}{\tilde\alpha^2}\,,
\end{equation}
and then Eq.~(\ref{eq:sig}) gives
\begin{equation}
 \sigma(\phi)\approx\frac{\alpha}{\tilde\alpha}\phi \, .
 \label{eq:sig_small}
\end{equation}
Although, as we discussed in the end of the previous subsection, the
explicit form of $V(\sigma)$ cannot be obtained, since it requires
the inversion of the  transcendental equation  (\ref{eq:sig}) of
$\sigma(\phi)$, its asymptotic form can be easily extracted, since
now  $\sigma(\phi)$ takes the simple form (\ref{eq:sig_small}) which
can be trivially inverted. In particular, for small $\chi$ the
potential becomes
\begin{equation}
 V_s(\sigma)\approx V_{s0}e^{-\tilde\alpha\sigma/\Mpl} \, .
 \label{eq:pot_small_chi}
\end{equation}
which for small slope can facilitate slow roll which can continue
for large values of $\chi$.
 Similarly, for very large values of $\chi$   ($\chi\gg m$),   Eqs.
(\ref{eq:K}),(\ref{kphi}) lead to
\begin{eqnarray}
\label{k=1}
 k^2(\phi)\approx1 \, ,
\end{eqnarray}
and then Eq.~(\ref{eq:sig}) gives
\begin{eqnarray}
 \sig\approx\phi-\frac{2}{\t\al}\ln\(\frac{\mu_m}{2}\)+\frac{2}{\al}
\ln\(\frac{ \t\al\mu_m}{\al+\t\al}\)  \, .
 \label{eq:sig_large}
\end{eqnarray}
Thus, for large $\chi$ the potential reads
\begin{eqnarray}
 V_l(\sig)\approx V_{l0}e^{-\al\sig/\Mpl} \, .
 \label{eq:pot_large_chi}
\end{eqnarray}
which gives rise to scaling solution for $\alpha>\sqrt{3}$, we shall
take $\alpha\simeq 10$ to satisfy the  nucleosynthesis constraint.

\begin{figure}[h]
\centering
\includegraphics[scale=.67]{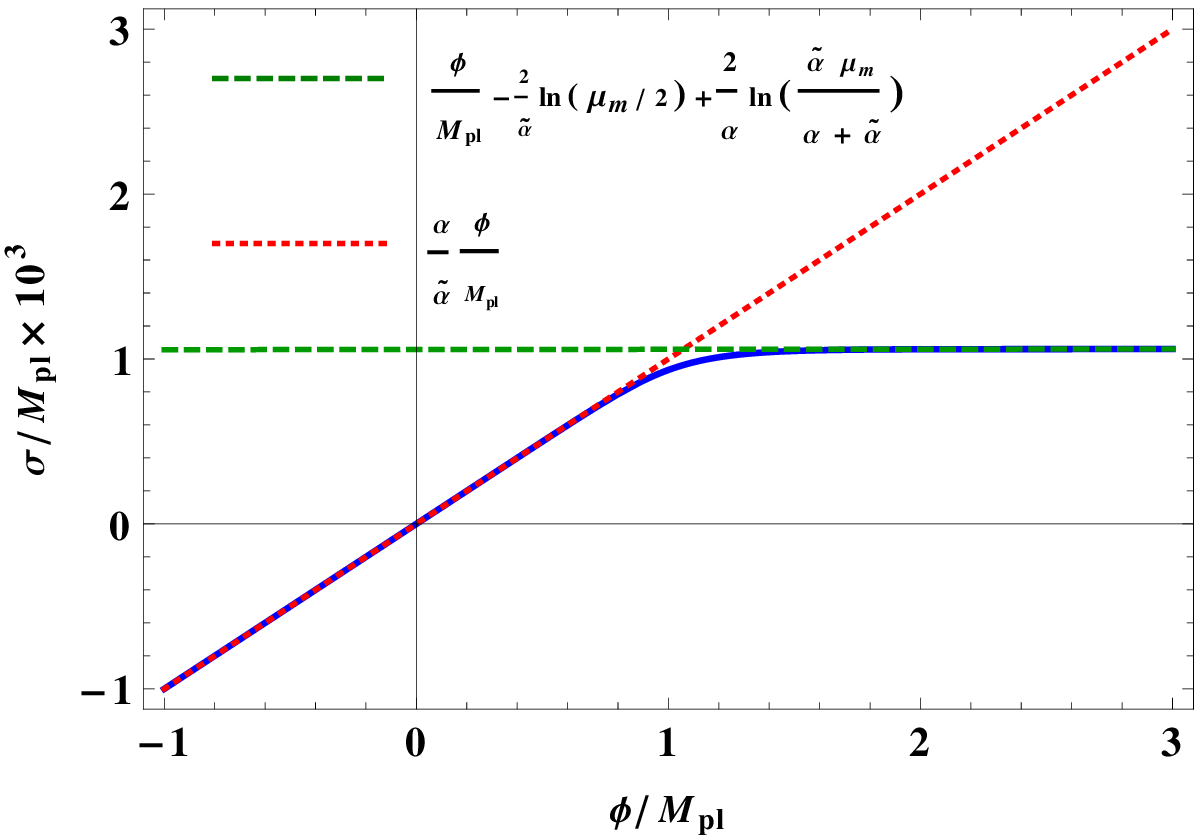}
\caption{Blue (solid) line represents the behavior of $\sig$ field (Eq.~(\ref{eq:sig})).
Red (dotted) line represents the Eq.~(\ref{eq:sig_small}) and the Green (dashed) line represents
the Eq.~(\ref{eq:sig_large}). The figure clearly shows the transition of $\sig$
field from Eq.~(\ref{eq:sig_small}) to Eq.~(\ref{eq:sig_large}). To plot
this figure
we have taken $\al=10$, $\t\al=0.01$ and $\mu_m=0.01$. If one changes the
value of $\al$ and $\t\al$ maintaining $\al>\t\al$ then only the transition
point changes
but the behavior remains the same. This plot can be extrapolated for small and
large values of $\phi$ field and nature remains the same. If we take values
$\t\al>\al$ then also the nature remains the same but the slopes of the straight
lines get changed.}
\label{fig:sig}
\end{figure}

From the above asymptotic expressions, we deduce that the
behavior of the canonical field
$\sig$ with respect to the non-canonical field
$\phi$, changes from a straight line with slope $\al/\t\al$ (for small
$\phi$) to a
straight line with slope 1, and $y$ axis intercepts at
$-\frac{2}{\t\al}\ln\(\frac{\mu_m}{2}\)+
\frac{2}{\al}\ln\(\frac{\t\al\mu_m}{\al+\t\al}\right)$
(for large $\phi$). This behavior is always true as long as $\al > \t\al$ and
 $\al > \mu_m$.
In Fig.~\ref{fig:sig} we present the change in $\sig$-field behavior, in
terms of the
$\phi$-field.
We next investigate the dynamics of unification in detail which
includes inflationary phase,thermal history and late time cosmic
acceleration. We shall also examine the issues related to relic
gravity waves, a generic feature of the scenario under
consideration. To this effect, we shall invoke the instant
preheating to circumvent the excessive production of gravity waves.

\section{Inflation}
\label{Inflation}

Having presented the scenario of variable gravity
\cite{Wetterich:2013jsa} in the Jordan and Einstein frames, in this
section we proceed to a detailed investigation of the inflationary
stage. As discussed earlier, at early times or equivalently for
negative values/small positive values of the field, the potential
$V(\phi)$ given by Eq.~(\ref{eq:pot}) reduces to the canonical
potential $V_s(\sigma)$ of (\ref{eq:pot_small_chi}), which
facilitates slow roll for small values of $\tilde{\alpha}\lesssim
\sqrt{2}$, where consistency with observations demands that
$\tilde{\alpha}\ll 1$. On the other hand, for very large values of
$\phi$, where $k(\phi)\to 1$ and the potential is given by
(\ref{eq:pot_large_chi}), we obtain the required scaling behavior in
radiation and matter era, for $\alpha \gtrsim 10$.

The $\sigma$-field slow
roll parameters, can be easily expressed in terms of $\phi$ as
\begin{eqnarray}
\label{eps1}
\epsilon&=&\frac{\Mpl^2}{2}\(\frac{1}{V}\frac{{\rm d}V}{{\rm d}\sig}\)^2
=\frac{\Mpl^2}{2k^2(\phi)}\(\frac{1}{V}\frac{{\rm d}V}{{\rm d}\phi}\)^2
=\frac{\alpha^2}{2k^2(\phi)}\, ,~~~~\,\,\\
\eta &=& \frac{\Mpl^2}{V}\frac{{\rm d^2}V}{{\rm d}\sig^2}
= 2\epsilon-\frac{\Mpl}{\alpha}\frac{{\rm d}\ep(\phi)}{{\rm d}\phi}\ ,
\label{eps2}
\end{eqnarray}
where we have made use of (\ref{dVsigma}). Since
$\alpha^2,\tilde{\alpha}^{-2}\gg 1$, the slow-roll regime lasts for
large values of $\phi$ (since $k^2=\alpha^2/2$) such that $X\equiv
\mu^2/m^2e^{\alpha\phi/M_p} \gg 1$, and thereafter the field crosses
to the kinetic regime where $k\simeq 1$.

Clearly, the
large-field slow-roll regime is of great physical interest. In this case the
slow-roll parameters are simplified to
\begin{eqnarray}
\epsilon=\eta=\frac{\tilde{\alpha}^2}{2}X  \ \to\ X_{\rm
end}=\frac{2}{\tilde{\alpha}^2}
\label{eq:X_end}
\end{eqnarray}
and the kinetic function is given by
\begin{equation}
\label{keq1}
k^2(\phi)\simeq \frac{\alpha^2}{\tilde{\alpha}^2 X}\ \to\
k_{\rm end}\simeq
\frac{\alpha}{\sqrt{2}}.
\end{equation}
We mention that $k^2$ interpolates between
$\alpha/\tilde{\alpha}$ and $1$, as the field evolves
from early epochs to late times. At the end of inflation $k_{\rm end} \simeq
6$, and then it quickly relaxes to $ k=1$ marking the beginning of the
kinetic regime. This transition takes place very fast, since the kinetic
function decreases exponentially with the field.

It is convenient to express the physical quantities in terms of the
non-canonical field $\phi$ too. It is then straightforward to write down the
Friedman equation in slow-roll regime as
\begin{equation}
H^2=\frac{\Mpl^2}{3}e^{-\alpha
\phi/\Mpl}\equiv\frac{\mu^2}{3m^2}\frac{\Mpl^2}{X},
\label{eq:Fried_end_in}
\end{equation}
which we shall use in the following discussion.

The Number of efoldings are given by,
\begin{eqnarray}
 \N(\phi) &=& \frac{1}{\al\Mpl}\int_\phi^{\phi_{\rm end}} k^2(\phi) {\rm d}\phi' \, , \nn \\
  &=& \frac{\al\(\phi_{\rm end}-\phi\)}{\t\al^2} \nn \\ && +\(\frac{1}{\al^2}-\frac{1}{\t\al^2}\)
  \ln \(\frac{m^2+\mu^2e^{\al\phi_{\rm end}/\Mpl}}{m^2+\mu^2e^{\al\phi/\Mpl}}\)\, ,~~~~~
  \label{eq:efold}
\end{eqnarray}
Where $\phi{\rm end}$ is the value of $\phi$ field at the end of inflation.
Now from Eq.~(\ref{eq:X_end}) we have $e^{\al\phi_{\rm end}/\Mpl}=2m^2/(\t\al^2\mu^2)$
which approximates the Eq.~(\ref{eq:efold}) by neglecting $\al^{-2}$ term
with respect to $\t\al^{-2}$,
\begin{eqnarray}
 \N(\phi) &\approx& \frac{1}{\t\al^2}\Bigg[\ln\(1+X^{-1}\)-\ln\(1+\frac{\t\al^2}{2}\)\Bigg]  .~~~~~
 \label{eq:efold1}
\end{eqnarray}
Fo given efoldings from Eq.~(\ref{eq:efold1}) we can calculate the value of $\phi$ when inflation
started.

The number of e-foldings in the large-$X$ approximation are given by
\begin{eqnarray}
\mathcal{N}(\phi_{\rm in})\simeq \frac{1}{\tilde{\alpha}^2X_{\rm in}} \, ,
\end{eqnarray}
where $\phi_{\rm in}$ designates the field-value where inflation commences.
The COBE normalized value of density perturbations \cite{Bunn:1996py,Bunn:1996da}
\begin{equation}
\delta^2_H=\frac{1}{150 \pi^2}\frac{1}{\Mpl^4}\frac{V_{\rm in}}{\epsilon}
\simeq 2\times 10^{-10} \, ,
\end{equation}
then allows us to estimate $V_{\rm in}$ as well as the important ratio of
parameters,
$\tilde{\alpha}^2 \mu^2/m^2$ in terms of the number of e-foldings, namely
\begin{eqnarray}
&&\frac{\tilde{\alpha}^2 \mu^2}{m^2}
=\frac{150\times \pi^2\times 10^{-10}}{\mathcal{N}^2}\\
&&V_{\rm in}=\mathcal{N}\frac{\tilde{\alpha}^2 \mu^2}{m^2}\Mpl^4
=\frac{150\times \pi^2\times 10^{-10}}{\mathcal{N}}\Mpl^4 \, .
\end{eqnarray}
Let us also note the important relationship between $H_{\rm in}$ and
$H_{\rm end}$ using the expressions of $X_{\rm in}$ and $X_{\rm end}$:
\begin{equation}
\frac{H^2_{\rm end}}{H^2_{\rm in}}=\frac{V_{\rm end}}{V_{\rm in}}
=\frac{X_{\rm in}}{X_{\rm end}}=\frac{1}{2\mathcal{N}}\, ,
\end{equation}
which in particular can be used to estimate the Hubble parameter at the end
of  inflation:
\begin{equation}
H^2_{\rm end}=\frac{\Mpl^2}{6}\frac{\tilde{\alpha}^2\mu^2}{m^2}=
\frac{25\pi^2\times 10^{-10}}{\mathcal{N}^2}\Mpl^2 \, .
\end{equation}

As mentioned in the introduction, the scenario under consideration
does not belong to the class of oscillatory models. In this case we
need to look for an alternative reheating mechanism, and a possible
candidate is the gravitational particle production
\cite{Ford:1986sy,Spokoiny:1993kt}. The space time geometry
undergoes a crucial transition at the end of inflation, involving
essentially a non-adiabatic process that gives rise to particle
production. Assuming thermalization of the so produced energy, the
energy density of radiation produced in this process at the end of
inflation is given by
\begin{equation}
\rho_{\rm rad} \simeq 0.01\times g_p H_{\rm end}^4 \, ,
\label{eq:rad_end}
\end{equation}
where $g_p$ is the number of different species produced
at the end of inflation, varying typically between 10
and 100. Thus, assuming $g_p\sim 100$, we obtain the radiation temperature
\begin{equation}
 T_{\rm end}\simeq 1.5 \times \frac{10^{-4}}{\mathcal{N}}\Mpl \, .
\end{equation}

Up to now we have kept the number of e-foldings arbitrary. This
number typically depends upon the reheating temperature and also the
scale of inflation. It can be estimated by considering a typical
length scale which leaves the Hubble scale during inflation at
$a=a_{\rm in}$ and re-enters the horizon today:
\begin{eqnarray}
&&k=a_{\rm in}H_{\rm in}=a_0H_0 \to \frac{k}{a_0 H_0}
=\frac{a_{\rm in}}{a_{\rm end}}\frac{a_{\rm end}}{a_0}\frac{H_{\rm in}}{H_0}
\nn \\
&&=e^{-\mathcal{N}}\frac{T_0}{T_{\rm end}}\frac{H_{\rm in}}{H_0} \, ,
\end{eqnarray}
which gives $\mathcal{N}\simeq 70$. Therefore,  the temperature at
the end of inflation is given by
\begin{eqnarray}
T_{\rm end} \simeq 3.6 \times 10^{12} \rm GeV.
\end{eqnarray}

We  then estimate the spectral index $n_s$ and the ratio of
tensor-to-scalar perturbations $r$ as

\begin{eqnarray}
&&n_s\approx 1-6\epsilon+2\eta=1-\frac{2}{\mathcal{N}}\simeq 0.97 \, ,
\label{eq:ns}\\
&&r\approx 16\epsilon=\frac{8}{\mathcal{N}}\simeq 0.11 \, .
\label{eq:r}
\end{eqnarray}
Eq.~(\ref{eq:ns}) and Eq.~(\ref{eq:r}) can be combined in a single equation,
namely
\begin{eqnarray}
 r=4\(1-n_s\) \, .
 \label{eq:ns_r}
\end{eqnarray}

\begin{figure}[h]
\centering
\includegraphics[scale=.45]{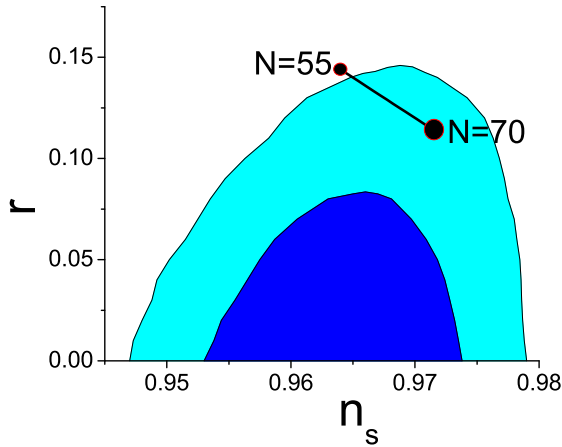}
\caption{1$\sig$ (blue) and 2$\sig$ (cyan) contours for
$Planck+WP+BAO$ data are shown on $n_s-r$ plane. We have also shown
the possible positions of $n_s$ and $r$ for the e-foldings
($\mathcal{N}$) 55 to 70 for the model under consideration. The
point for $\mathcal{N}=55$ on the $n_s-r$ plane is outside the
2$\sig$ contour but slight higher values of the e-foldings result
points within the 2$\sig$ contour.} \label{fig:t_to_s}
\end{figure}

In Fig.~\ref{fig:t_to_s} we present the $68\%$ and $95\%$ contours
on $n_s-r$ plane, using the data of $Planck+WP+BAO$
\cite{Ade:2013uln}. On top of them we depict the $n_s$ and $r$
values calculated in our model using (\ref{eq:ns}) and (\ref{eq:r})
respectively, having considered the e-foldings
 $(\mathcal{N})$ between 55 and 70. It seems that the value $\mathcal{N}=55$
is ruled out upto  2$\sig$ level for this model. But the values slightly
higher than 55 are well within the 2$\sig$ level. The line shown in the Fig.
\ref{fig:t_to_s} follows   Eq.~(\ref{eq:ns_r}).
In the subsection to follow, we consider the problem related to
excessive production of relic gravity waves.
\subsection{Relic gravity waves and nucleo-synthesis constraint on reheating
temperature}
\label{RGW}

Let us assume that gravitational particle production is
the sole mechanism for reheating
\cite{Kofman:1994rk,Dolgov:1982th,Abbott:1982hn,
Ford:1986sy,Spokoiny:1993kt,Kofman:1997yn,Shtanov:1994ce,Campos:2002yk}.
In this case using Eqs.~(\ref{eq:Fried_end_in}) and (\ref{eq:rad_end})
and considering $\mathcal{N}=70$ we have,
\begin{equation}
\label{gp}
 \left(\frac{\rho_\phi}{\rho_{r}}\right)_{\rm end}\simeq
\frac{\Mpl^2 H^2_{\rm end}}{0.01\times g_p H^4_{\rm end}}\simeq 2\times
10^{11}; (g_p\simeq 100) \, .
\end{equation}

This ratio is typically $10^{16}$ in braneworld models
\cite{Sahni:2001qp,Sami:2004xk}. Hence, it
takes long for the radiative regime to commence. In those models
similar to our situation potential is very steep after inflation and
therefore $\rho_\phi\sim 1/a^6$ till radiation takes over. During
this regime called kinetic regime the gravity wave(produced during
inflation)  amplitude enhances and violates nucleosynthesis
constraints at the commencement of the radiative regime
\cite{Sahni:2001qp,Sami:2004xk}. We have to
check it here also.

The quantum mechanical production of gravity waves during inflation
is a generic feature of the scenario.
The tensor perturbations
$h_{ij}$ satisfy the Klein-Gordon equation $ \Box h_{ij}=0$
\cite{Grishchuk:1974ny,Grishchuk:1977zz}
which gives,
\begin{equation}
\label{KG}
 \ddot{\varphi}_k(\tau)+2\frac{\dot a}
{a}\varphi_k(\tau)+k^2\varphi_k(\tau)=0 \, ,
\end{equation}
where $h_{ij}\sim \varphi_ke^{ikx}e_{ij}$( $e_{ij}$ is polarization
tensor); $\tau$ (${\rm d}\tau={\rm d}t/a$) is conformal time and $k$ is comoving
wave number. As pointed out in the preceding discussion, inflation
is approximately exponential thereby $a=\tau_0/\tau$ and
$H_{\rm in}=-1/\tau_0$ is the Hubble parameter during inflation. The
``in" state $\varphi_{\rm in}^{(+)}(k,\tau)$ corresponds to the positive
frequency solution of Eq.~(\ref{KG}) in the adiabatic
vacuum($\varphi_{\rm in}^{(+)}(k,\tau)=(\pi
\tau_0/4)^{1/2}(\tau/\tau_0)^{3/2}H^{(2)}_{3/2}(k\tau) $. After
inflation has ended, universe from quasi de Sitter phase makes
transition to the phase characterized by power law expansion. In the
standard scenario, the post inflationary evolution is described by
radiative regime whereas in the quintessential inflation, the after
inflation transition is to kinetic phase with stiff equation of
state parameter \cite{Sahni:2001qp,Sami:2004xk}.
This transition involves non-adiabatic change of
geometry. We shall assume that post inflationary dynamics is
described by power law expansion, $a=(t/t_0)^p\equiv
(\tau/\tau_0)^{1/2-\mu}$ where $ \mu\equiv
3/2((w-1)/(3w+1))$ with $w$ being the post
inflationary equation of state parameter. Let us notice that $\mu=0$
in kinetic regime($w=1$).
 The "out" state contains
both positive and negative frequency solutions to (\ref{KG}),
\begin{equation}
\varphi_{\rm out}=\alpha \varphi_{\rm out}^{(+)}+\beta \varphi_{\rm out}^{(-)} \, ,
\end{equation}
where $\alpha$ and $\beta$ are Bogoliubov coefficients
\cite{Sahni:2001qp}. The "out" is state is given by,
$\varphi^{(+,-)}_{\rm out} =(\pi \tau_0/4)^{1/2}(\tau/\tau_0)^\mu
H^{(2,1)}_{|\mu|}(k\tau) $. The energy density of relic gravity
waves depends upon $\beta$ \cite{Sahni:2001qp,Sahni:1990tx},
\begin{equation}
\label{rhog}
 \rho_g=<T_{00}>=\frac{1}{\pi^2a^2}\int{{\rm d}kk^3|\beta|^2} \, .
\end{equation}
During kinetic regime, $|\beta_{\rm kin}|^2\sim (k\tau_{\rm kin})^{-3}$, as a
result using (\ref{rhog}), we obtain,
\begin{equation}
\label{rhog2}
\rho_g=\frac{32}{3\pi}h^2_{\rm GW}\rho_b\left(\frac{\tau}{\tau_{\rm kin}}\right)
\end{equation}
where $\rho_b$ is the background energy density made by radiation
and scalar stiff matter. While deriving (\ref{rhog2}), we made use
of the fact that $H_{\rm in}=-1/\tau_0$. Since, at radiation, field
equality ($\tau=\tau_{\rm eq}$), $\tau_{\rm eq}/
\tau_{\rm kin}=(T_{\rm kin}/T_{\rm eq})^2$ and $\rho_b=2 \rho_{r}$, we have from Eq.~(\ref{rhog2}),
\begin{equation}
\label{rhogr}
\left(\frac{\rho_g}{\rho_{r}}\right)_{\rm eq}=\frac{64}{3
\pi}h^2_{\rm GW}\left(\frac{T_{\rm end}}{T_{\rm eq}}\right)^2 \, ,
\end{equation}
where $h_{\rm GW}$ is the dimensionless gravity amplitude
which needs to be fixed in each model imposing COBE normalization
\cite{Bunn:1996da,Bunn:1996py},
\begin{eqnarray}
\label{hgw}
 h^2_{\rm GW}&=&\frac{H^2_{\rm in}}{8 \pi \Mpl^2}=\frac{\mathcal{N}}{24
\pi}\left(\frac{\tilde{\alpha}^2\mu^2}{m^2}\right)=\frac
{\mathcal{N}}{4 \pi}\frac{H^2_{\rm end}}{\Mpl^2} \nn \\ && \simeq 2.8\times 10^{-11} \, .
\end{eqnarray}
Let us notice from (\ref{rhogr}) that longer is the kinetic regime
smaller would be $T{\rm eq}$ thereby larger would be the ratio of energy
densities of relic gravity waves and radiation at equality. It may
also be worthwhile to note from (\ref{rhog}) that $\rho_g\sim 1/a^4$
for $\omega>1/3$ whereas $\rho_g\sim \rho_b$ if $\omega<1/3$ and
during radiation era also $\rho_g$ approximately tracks the
background. It is the specific behavior of $\rho_g$ during kinetic
regime which causes problem.

\begin{figure}[h]
\centering
\includegraphics[scale=.8]{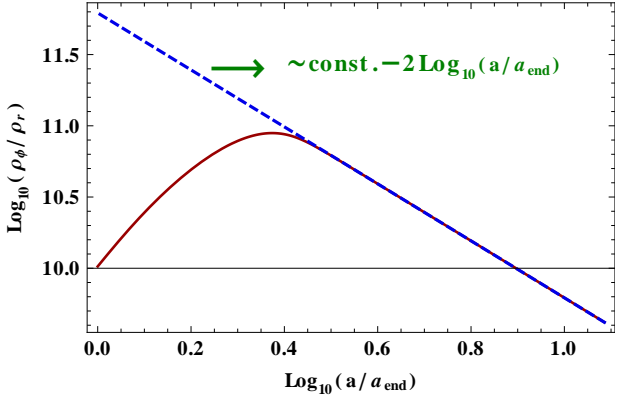}
\caption{Post inflationary evolution of $\rho_\phi$ is shown from
the end of inflation, $a_{\rm end}=1$ on logarithmic scale. The blue
dotted straight line corresponds to $const/a^2$ on the log scale. It
touches the curve $\rho_\phi/\rho_r$ around $0.4$ on the $x$ axis
which signals the commencement of kinetic regime. Figure shows that
kinetic regime establishes pretty past after inflation ends.}
\label{fig:kin}
\end{figure}

 For simplicity we shall here made an
approximation that the field after inflation instantaneously come to
kinetic regime($\rho_\phi\sim 1/a^6$). In fact, $\rho_\phi\sim
1/a^2$ at $\phi=\phi_{\rm end}$ and soon thereafter, the field enters
the kinetic regime which happens pretty fast as potential is steep.
Thus we shall assume that $H_{\rm end}\simeq H_{\rm kin}$ and $T_{\rm end}\simeq
T_{\rm kin}$. Numerical calculations show that kinetic regime fast
commences, (see Fig.~\ref{fig:kin}) after the end of inflation; our
estimates do not change significantly
 by adopting the said approximation.
Since $T_{\rm eq}\sim T_{\rm end}/a_{\rm eq}$, we have,
\begin{equation}
\left(\frac{\rho_\phi}{\rho_{r}}\right)_{\rm end}=\left(\frac{T_{\rm end}}{T_{\rm eq}}\right)^2
\end{equation}
we have the following relation,
\begin{equation}
\left(\frac{\rho_\phi}{\rho_{r}}\right)_{\rm end}=\frac{3\pi}{64}
\left(\frac{\rho_g}{\rho_{r}}\right)_{\rm eq}\frac{1}{h^{2}_{\rm GW}} \, .
\end{equation}
 As for
$\rho_g/\rho_{r}$ at equality, nucleosynthesis dictates that it
should be less than $0.2$ \cite{Sahni:2001qp}. We know left hand side, so if we estimate
gravity wave amplitude, we can find out whether the gravitational
particle production can do the job. Indeed, we find using
Eq.(\ref{hgw}),
\begin{equation}
\label{nuc}
 \left(\frac{\rho_\phi}{\rho_{r}}\right)_{\rm end}\lesssim
\frac{3\pi}{64}\times 0.2 \times \frac{4\pi}{
\N}\frac{\Mpl^2}{H^2_{\rm end}}\simeq 10^{9} \, .
\end{equation}
Comparing the estimate with the one obtained by using  (\ref{gp}),
we conclude that even if we take $g_p \sim 100$, the gravitational
particle production does not meet the requirement imposed by the
nucleosynthesis constraint at the commencement of radiative regime.
It should also be noted that kinetic regime does not set
instantaneously; incorporating evolution from the end of inflation
to the beginning of kinetic regime further worsens the situation.
Gravitational particle production is clearly an inefficient process
and we should therefore look for an alternative way of reheating.
Instant preheating provides with an efficient mechanism which suits
to the quintessential inflation scenario under consideration.

Let us also quote the spectral energy density of the gravitational
wave (see Ref.\cite{Sahni:2001qp} for details),
\begin{equation}
 \Omega_{\rm GW}(\lam)=\frac{1}{\rho_c} \frac{{\rm d}\rho_g}{{\rm d}\ln k} \, ,
\end{equation}
where $\rho_c$ is the critical energy density.

In different epochs, the form of $\Omega_{\rm GW}$ is given by,
\begin{eqnarray}
&&\Omega_{\rm GW}^{\rm (MD)}= \frac{3}{8\pi^3}h_{\rm GW}^2
\Omega_{m0}\(\frac{\lam}{\lam_h}\)^2 \, ,  \lam_{\rm MD}<\lam\leq \lam_h  ,~~~~~~\\
&&\Om_{\rm GW}^{\rm (RD)}(\lam)=\frac{1}{6\pi}h_{\rm GW}^2\Omega_{r0},
~~~~~~~~~ \lam_{\rm RD}<\lam\leq\lam_{\rm MD} \, ,~~~~~ \\
&&\Om_{\rm GW}^{\rm (kin)}(\lam)=\Om_{\rm GW}^{\rm (RD)}\(\frac{\lam_{\rm RD}}{\lam}\) \, ,
~~~~~  \lam_{\rm kin}<\lam\leq\lam_{\rm RD} \, ,~~~~~
\end{eqnarray}
where,
\begin{eqnarray}
 \lam_h &=& 2cH_0^{-1}\, , \\
 \lam_{\rm MD} &=& \frac{2\pi}{3}\lam_h\(\frac{\Om_{r0}}{\Om_{m0}}\)^{1/2} \, , \\
 \lam_{\rm RD}&=& 4\lam_h\(\frac{\Om_{m0}}{\Om_{r0}}\)^{1/2}\frac{T_{\rm MD}}{T_{\rm rh}} \, ,\\
 \lam{\rm kin} &=& cH_{\rm kin}^{-1}\(\frac{T_{\rm rh}}{T_0}\)\(\frac{H_{\rm kin}}{H_{\rm rh}}\)^{1/3} \, ,
\end{eqnarray}
where ``MD'', ``RD'' and ``kin'' represent matter dominated, radiation dominated and
kinetic energy dominated epochs and $\lam$ represents wavelength. $H_0$ is the present
value of Hubble parameter and $\Om_{m0}$ and
$\Om_{r0}$ are the present values of matter and radiation energy densities.
$T_{\rm rh}$ and $H_{\rm rh}$
are the reheating temperature and Hubble parameter respectively which are approximately same as
the temperature and Hubble parameter at the end of the inflation respectively.

\subsection{Instant preheating}
\label{Instant Preheating}

In this subsection, we shall describe instant preheating applied to
the scenario under consideration. We shall demonstrate its viability
to tackle the problem associated with relic gravity waves.

 Inflation ends when
$\phi=\phi_{\rm end}$ which, for convenience, we can shift to the origin
by translating the field, $\phi'=\phi-\phi_{\rm end}$ without the loss
of generality. In what follows we would keep using $\phi$
remembering that the translated field $\phi<0$. We next assume that
$\phi$ interacts with a new field $\chi$ which interacts with a
Fermi field via Yukawa interaction,
\begin{equation}
\mathcal{L}_{\rm int}=-\frac{1}{2}g^2\phi^2\chi^2-h\bar{\psi}\psi\chi
\end{equation}
where $\chi$ does not have bare mass, its effective mass is given
by, $m_{\chi}=g|\phi|$ and couplings $g$ $\&$ $h$ are assumed to be
positive. In the model under consideration, the field $\phi$ soon
comes to kinetic regime after inflation has ended as potential is
steep there. In this case the production of $\chi$  particles may
commence provided $m_{\chi}$ changes non-adiabatically
\cite{Felder:1998vq,Felder:1999pv},
\begin{equation}
\label{nad}
 \dot{m}_{\chi}\gtrsim  m^2_{\chi} \to \dot{\phi}\gtrsim g\phi^2 \, .
\end{equation}
The condition for particle production (\ref{nad}) can be satisfied
provided,
\begin{equation}
|\phi|\lesssim |\phi_p|=\sqrt{\frac{\dot{\phi}_{\rm end}}{g}} \, .
\label{eq:phi_p}
\end{equation}
Using slow roll equations (\ref{eps1}) and (\ref{eps2}), it can be noticed that,
\begin{equation}
\dot{\phi}_{\rm end}=\frac{\al}{k_{\rm end}^2}\sqrt{\frac{V_{\rm end}}{3}};~~k_{\rm end}=\frac{\alpha}{\sqrt{2}} \, .
\end{equation}
Since $\phi_p \lesssim \Mpl$, from Eq.~(\ref{eq:phi_p}) we have a constraint on the coupling
$g$,
\begin{equation}
\frac{\dot{\phi}_{\rm end}}{g}\lesssim \Mpl^2 \to g\gg
\frac{2}{\al\Mpl^2} \sqrt{\frac{V_{\rm end}}{3}} \, .
\end{equation}
Further, we can estimate the production time,
\begin{equation}
\delta{t}_p \sim
\frac{|\phi|}{\dot{\phi}}=g^{-1/2}\dot{\phi}_{\rm end}^{-1/2} \, .
\end{equation}
Using then uncertainty relation gives us the estimate for wave
number, $k_p\simeq \delta{t}^{-1}_p\simeq
\sqrt{g\dot{\phi}_{\rm end}}$. We then can find out the occupation
number for $\chi$ particles \cite{Felder:1999pv,Kofman:1997yn},
\begin{equation}
n_k\sim e^{-\pi k^2/k^2_p} \, ,
\end{equation}
which gives the number density of $\chi$ particle,
\begin{equation}
N_{\chi}=\frac{1}{(2\pi)^3}\int_0^\infty{n_k d^3{\bf
k}}=\frac{(g\dot{\phi}_{\rm end}) ^{3/2}}{(2\pi)^3} \, .
\end{equation}

The energy density of created particles $\chi$ is given by,
\begin{eqnarray}
\rho_{\chi}=N_\chi m_\chi=\frac{(g\dot{\phi}_{\rm end})^{3/2}}{(2\pi)^3}g|\phi_p|
=\frac{g^2V_{\rm end}}{6\pi^3\al^2} \, .
\label{eq:rho_chi_ph}
\end{eqnarray}
If the particle energy produced at the end of inflation is supposed
to be thermalized then using Eq.~(\ref{eq:Fried_end_in}) and
Eq.~(\ref{eq:rho_chi_ph}) we find,
\begin{equation}
\label{glim}
\left(\frac{\rho_\phi}{\rho_r}\right)_{\rm end}\simeq
\frac{6\pi^3\al^2}{g^2}
\end{equation}
Using (\ref{glim}), we can find lower limit on the coupling $g$ by
invoking the relic gravity constraint on $\rho_\phi/\rho_r$ from
(\ref{nuc}),
\begin{equation}
g \gtrsim  6\al\times 10^{-5}
\end{equation}

 Let us further note that
\begin{eqnarray}
&&\delta t_p H_{\rm end} \simeq
\sqrt{\frac{\al}{2g\Mpl^2}}\(\frac{V_{\rm end}}{3}\)^{1/4}< \frac{4.5\times 10^{-5/2}}{\N} \nn \\
&&\Rightarrow
\delta t_p\ll H^{-1}_{\rm end} \, ,
\end{eqnarray}
since $\N\sim 70$. This tells us that during particle production expansion can be
ignored.
 Let us also notice that $\phi_p\simeq 4\times 10^{-4}$ which
implies that particle production takes place all most
instantaneously after inflation has ended.
 Since $\phi$ runs fast after inflation has ended, the mass of
$\chi$ grows larger making it to decay into $\bar{\psi}\psi$, the
decay width is given by
\begin{equation}
\Gamma_{\bar{\psi}\psi}=\frac{h^2
m_\chi}{8\pi}=\frac{h^2}{8\pi}g|\phi|
\end{equation}

\begin{figure}[h]
\centering
\includegraphics[scale=.7]{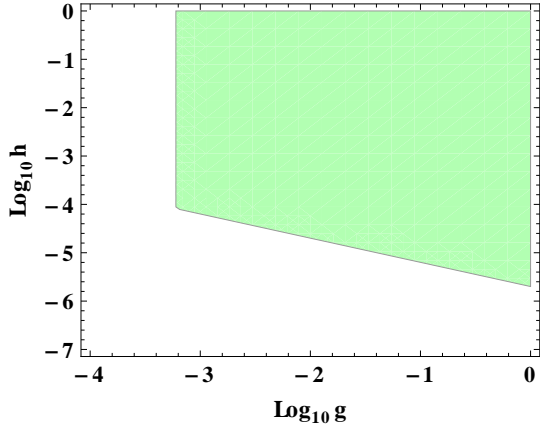}
\caption{Figure depicts the parameter space of ($g,h$). Shaded
region shows the allowed values of the parameters where preheating
is efficient. $\al$ is considered to be 10.} \label{fig:gh}
\end{figure}

 We should now worry about
the  back reaction of $\chi$ on the post inflationary dynamics of
$\phi$. Around $\phi=0$ where inflation ends, $\rho_\phi \sim 1/a^2$
thereby the field potential as well as the dissipative term in the
evolution equation for $\phi$ evolve slower than $\rho_\chi$. On the
other hand, the decay rate is larger of larger values of $\phi$ as
$m_\chi$ gets larger. Hence, the decay of $\chi$ into Fermions would
be accomplished before the back reaction of $\chi$ on $\phi$
evolution becomes important provided that,
\begin{equation}
\Gamma_{\bar{\psi}\psi}\gg H_{\rm end}\to h^2 \gtrsim 8\pi
\frac{H_{\rm end}}{g|\phi|}
\end{equation}
Since $\phi\lesssim \Mpl$, the above estimate implies that $h \gtrsim
2g^{-1/2}\times 10^{-6}$ which gives the lower bound on the
numerical value of the coupling $h$. Fig.~\ref{fig:gh} shows the
allowed values of $g$ and $h$. It is clear that is a wide region in
the parameter space where the instant preheating is quite efficient.

\begin{figure}[h]
\centering
\includegraphics[scale=.75]{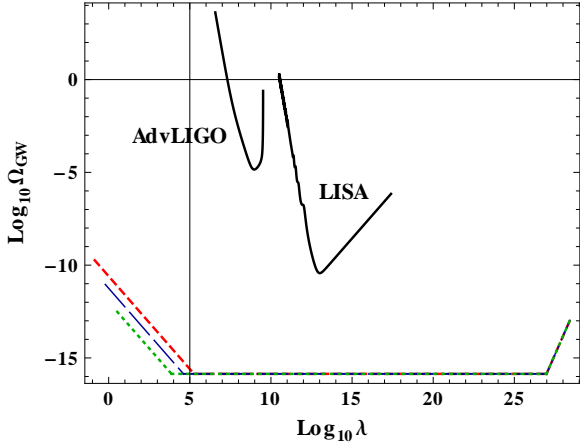}
\caption{Spectral energy density of relic gravity wave background
for different reheating temperatures. Red (dashed), Blue (long dashed)
and Green (dotted) lines for $g=5\times 10^{-4}$, $0.01$ and $0.3$
respectively. $\al$ is taken to be $10$. Also we have considered
$\mathcal{N}=70$ for this plot but it is checked that the behavior
does not change significantly for the variation of $\mathcal{N}$ from
50 to 70. Black solid curves represent the
expected sensitivity curves of Advanced LIGO and LISA.}
\label{fig:RGW}
\end{figure}

Fig.~\ref{fig:RGW} shows the spectral energy density ($\Om_{\rm GW}$)
of relic gravitational wave background along with the sensitivity curve
of AdvLIGO \cite{LIGO,aLIGO} and LISA \cite{LISA,LISA1}.
To plot Fig.~\ref{fig:RGW} we have taken the present
values of matter and radiation energy density to be 0.3 and
$9\times 10^{-5}$ respectively.

Next, we turn to late time dynamics of the model.

\section{Late time Cosmology: Dark Energy}
\label{Late times evolution}

In this section we investigate the cosmological behavior at late times,
where as we mentioned in the Introduction the scenario at hand leads to an
effective dark energy driving universe acceleration.

\subsection{Evolution equations}
\label{Cosmological equations}

We consider the spatially flat  Friedmann-Robertson-Walker (FRW)
cosmology,
\begin{align}
\label{metric}
 ds^2 = -N^2dt^2 +a(t)^2\delta_{ij} dx^idx^j ~,
\end{align}
Varying the action (\ref{eq:action_E3}) with respect to the metric
$g_{\mu\nu}$ and setting $N=1$, we obtain the two Friedmann equations:
\begin{eqnarray}
 &&3H^2\Mpl^2 = \frac{1}{2}\dot\sig^2+V(\sig)+\rho_{m}+\rho_{r}+\rho_{\nu} \, ,
 \label{eq:Fried1} \\
 &&\(2\dot H+3H^2\)\Mpl^2 = -\frac{1}{2}\dot\sig^2+V(\sig)-\frac{1}{3}
\rho_{r}-p_{\nu}  , \, ~~~~~~
 \label{eq:Fried2}
\end{eqnarray}
where, as we mentioned, the neutrino pressure $p_{\nu}$ behaves as radiation
during the early times but it behaves like non-relativistic matter during the
late times. Varying the action (\ref{eq:action_E3}) with respect to the field
$\sig$ leads to its equation of motion
\footnote{Variation of $\S_\nu$ with
respect to $\sig$ reads as
\begin{eqnarray}
 \frac{1}{\sqrt{-g}}\frac{\delta\S_\nu}{\delta\sig}
 =\frac{1}{\sqrt{-g}}\frac{\delta\S_\nu}{\delta\phi}
\frac{\partial\phi}{\partial\sig}
 =\frac{\C_{,\phi}}{\C}\frac{T^{(\nu)}}{k(\phi)} \nn \, .
\end{eqnarray}
}:
\begin{eqnarray}
 \ddot\sig+3H\dot\sig=-\frac{{\rm d}V(\sigma)}{{\rm d}\sig}-\frac{\partial
\ln
m_\nu}{\partial\sig}\(\rho_\nu-3p_\nu\) \, .
 \label{eq:eom_sig}
\end{eqnarray}
Additionally, note that  relation (\ref{eq:mnu_phi}), using (\ref{sigmadef})
and  (\ref{eq:dsig}), gives
\begin{eqnarray}
 \frac{\partial \ln m_\nu}{\partial\sig} = \frac{\t\gam\al}{\Mpl k(\phi)} \, .
 \label{eq:mnu_sig}
\end{eqnarray}

Let us make an important comment here. During radiative regime, the
last term in the r.h.s. of Eq.~(\ref{eq:eom_sig}) does not
contribute as
 during that era neutrinos behave like radiation and the energy
momentum tensor is traceless.  On the contrary, at late times,
neutrinos behave as non-relativistic matter. As a result, the last
term in the r.h.s. of (\ref{eq:eom_sig}) is non-zero and the
non-minimal coupling between the scalar field and the neutrinos
builds up which plays a vital role in the model under consideration.

According to \cite{Wetterich:2013jsa},  as we mentioned in
(\ref{k=1}), during and after radiation era,  we can take $\chi\gg
m$ and $k(\phi)\approx 1$. Thus, the neutrino mass
(\ref{eq:mnu_phi}) at late times  exhibits an effective behavior
\begin{eqnarray}
 m_{\nu,\rm eff}(\sig)=m_{\nu,0}e^{\t\gam\al\sig/\Mpl} \, ,
\end{eqnarray}
which shows the same behavior as Eq.~(\ref{eq:mnu_phi}) that gives
rise to the same type of effective potential like Eq.
(\ref{eq:pot_eff_phi}). In this case, the neutrino conservation
equation (\ref{eq:neutcontefphi1}) effectively reads
\begin{eqnarray}
 \dot\rho_\nu+3H(\rho_\nu+p_\nu)=\frac{\t\gam\al} {\Mpl}\dot\sig
(\rho_\nu-3p_\nu) \, .
\end{eqnarray}

We stress here that in the scenario at hand, the late-time
acceleration is attributed to the combined effect of neutrinos and
scalar field, that is the effective dark energy sector includes
these two contributions, namely its energy density and pressure read
\begin{eqnarray}
\label{rhoDE}
 &&\rho_{\rm
DE}\equiv\rho_\nu+\rho_\sig=\rho_\nu+\frac{1}{2}\dot\sig^2+V(\sigma)
,\\
 &&p_{\rm DE}\equiv p_\nu+p_\sig=p_\nu+\frac{1}{2}\dot\sig^2-V(\sigma),
 \label{pDE}
\end{eqnarray}
and they obey the continuity equation
\begin{eqnarray}
\label{DEevol}
 \dot\rho_{\rm DE}+3H\(\rho_{\rm DE}+p_{\rm DE}\)=0 \, .
\end{eqnarray}

There is still one missing information in order for the above cosmological
equations to close, namely the behavior of the neutrino
equation-of-state parameter $w_\nu\equiv p_\nu/\rho_\nu$, which determines
the neutrino pressure $ p_\nu$ that enters into $p_{\rm DE}$, and then into
the conservation equation (\ref{DEevol}).

As we discussed in detail in section (\ref{EF}), before and during
the radiation era neutrinos are relativistic and behave as
radiation, while during and after the matter era neutrinos become
non-relativistic and $w_\nu$ becomes $0$. Thus, a complete and
detailed investigation of the thermal history of the universe
requires the exact behavior of $w_\nu$, that is its specific form
interpolating between these two regimes. Expressing the universe
evolution through the redshift $z$, for convenience, one can have
several $w_\nu(z)$ parameterizations with the above required
properties, namely, the interpolation of the equation of state
parameter between $1/3$ $\&$ $0$ \cite{Collodel:2012bp}. In this
work we desire to have a better control on the features of this
transition, namely, the epoch around which the  transition is
realized and the duration of realization. We shall use the following
ansatz for $w_\nu(z)$,
\begin{eqnarray}
 w_\nu(z)=\frac{p_\nu}{\rho_\nu}=\frac{1}{6}\left\{1+\tanh\left[\frac{
\ln(1+z)-z_{\rm eq} } {
z_{\rm dur}}\right]\right\} \, .
 \label{eq:w_nu}
\end{eqnarray}
In the above expression  $z_{\rm eq}$ determines the moment around
which the transition takes place; the choice   for the transition
redshift where  matter and radiation energy densities become equal
is reasonable. Additionally, $z_{\rm dur}$determines how fast this
transition is realized. In particular, having in mind that varying
mass, neutrinos become non-relativistic after their mass turns
constant \cite{Amendola:2007yx,Wetterich:2007kr}, and imposing the
physical requirement that the varying mass of neutrinos has to be
non-relativistic at the recent cosmological past, we deduce that we
need  a large  value of $z_{\rm dur}$ such that the transition is
smooth. However, the exact
 $z_{\rm dur}$-determination requires exact knowledge of the redshift
$z_{\rm NR}$ after
which neutrinos become non-relativistic, which according to
\cite{Amendola:2007yx,Wetterich:2007kr} is
$z_{\rm NR}\in (2-10)$ for $m_\nu\in (0.015-2.3)$, while according to
\cite{LaVacca:2012ir} it is  $z_{\rm NR}<4$.

Finally, in order to compare with observations, we introduce the
dimensionless density parameters for radiation,
matter, neutrinos and
scalar field, respectively as
\begin{eqnarray}
 \Omega_m &=& \frac{\rho_m}{3H^2\Mpl^2} \, ,
 \label{eq:Omega_m}\\
   \Omega_r &=& \frac{\rho_r}{3H^2\Mpl^2} \, ,
 \label{eq:Omega_r}\\
  \Omega_\nu &=& \frac{\rho_\nu}{3H^2\Mpl^2} \, ,
 \label{eq:Omega_nu}\\
  \Omega_\sig &=& \frac{\rho_\sig}{3H^2\Mpl^2} \, .
 \label{eq:Omega_sig},
\end{eqnarray}
 and thus, according to (\ref{rhoDE}),
 \begin{eqnarray}
 \Om_{\rm DE}=\Om_\sig+\Om_\nu \, .
 \label{eq:density_DE0}
\end{eqnarray}
Lastly, the equation-of-state parameters of the total matter content
in the universe  of the scalar-field sector and of the dark-energy
sector can be written as
\begin{eqnarray}
 w_{\rm eff} &=& -1-\frac{2}{3}\frac{\dot H}{H^2},
 \label{eq:w_eff0} \\
 w_{\sig} &=& \frac{p_\sigma}{\rho_\sigma} \, ,
 \label{eq:w_sig0} \\
 w_{\rm DE} &=& \frac{w_{\rm eff}-\frac{1}{3}\Om_r}{\Om_{\rm DE}}
\, .
 \label{eq:w_DE0}
\end{eqnarray}
In what follows we shall present our numerical results.

\subsection{Post inflationary dynamics: the epoch sequence}
\label{Transient Behavior}

Let us not examine the thermal history of the universe, that is we
are interesting in its transient behavior from inflation  to the
present epoch. Due to the complexities of the cosmological equations
of the previous subsection, no exact analytical solutions are
possible and one needs to perform a numerical elaboration. In
particular, we numerically evolve the cosmological equations
(\ref{eq:Fried1}), (\ref{eq:Fried2}),
(\ref{eq:eom_sig}),(\ref{eq:mnu_sig}),
(\ref{DEevol}),(\ref{eq:w_nu}), focusing on the evolution of
observables like the various density and equation-of-state
parameters. For the numerical evolution we consider $\al=10$,
$\t\gam=30$ and\footnote{ The parameter $\t\gam$ enters in the expression of the minimum of the effective potential (52) given by $V_{\rm eff,min}=\left(1+1/\t\gam\right)V_{\rm min}$ which tells us that $\t\gam\gg1$ for $V_{\rm eff,min}\sim H_0^2\Mpl^2$. $\t\gam$ also enters in the expression of equation of state parameter of dark energy whose value at the attractor point is given by $w_{\rm DE}=-\t\gam/(1+\t\gam)$. There is nothing special about $\t\gam=30$. It could be any large value such that $w_{\rm DE}$ falls within the observed value of the equation of state parameter(For instance, $\t\gam=30$. $w_{\rm DE}\approx0.97$).} $z_{\rm dur}=3.6$ and $10$, and for the initial conditions of
the radiation and scalar field we use the ratio of $\rho_r$ and
$\rho_\sig$ we obtain at the end of inflation. Additionally, for
matter and neutrinos we impose $\Om_{m0}\approx0.3$ and
$\Omega_{\nu0}\simeq 0.01$ at the present epoch. Finally, the value
of $z_{\rm dur}$ is set in order for the neutrinos to become
non-relativistic at the recent past (considering $z_{\rm NR}\sim
2-10$ \cite{Wetterich:2007kr}).

\begin{figure}[ht]
\centering
\includegraphics[scale=.7]{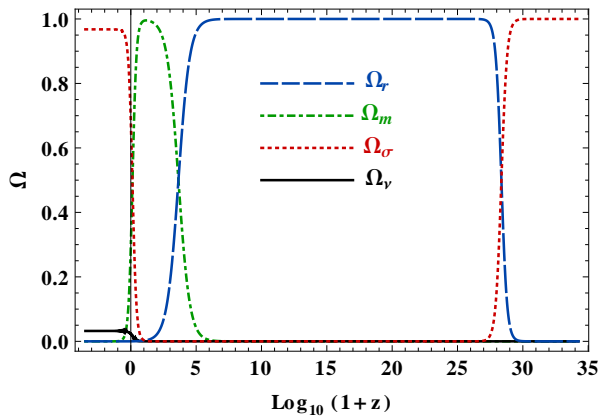}
\caption{Figure shows the evolution of different density parameters
($\Om$) are shown here. $\Om_r$ (Blue long dashed), $\Om_m$ (Green
dot-dashed), $\Om_\nu$ (Black solid), $\Om_\sig$ (Red dotted)
represent the density parameters for radiation, matter, neutrino and
scalar field $\sigma$
respectively.
This figure clearly shows the cosmological sequences starting from a
scalar field kinetic regime to late time dark energy dominated era.
We have used the numerical values, $\al=10$, $\tilde{\gamma}=30$ and
$z_{\rm dur}=3.6$ for plotting the figure. Since at the end of
inflation, $k_{\rm end}=\al/\sqrt{2}$ we taken the initial value of
$\lam\sim\mathcal{O}(1)$.} 
\label{fig:density}
\end{figure}

In Fig.~\ref{fig:density} we depict the evolution of  $\Om_m$, $\Om_r$,
$\Om_\nu$ and $\Om_\sig$.
The figure shows the evolution of universe from the kinetic regime
(where the scalar-field kinetic energy is dominant)  after the end
of inflation followed by the radiation and matter eras. Finally, the
universe enters into the dark-energy epoch and late-time
acceleration commences. Apart from the above standard thermal
history of the universe, which acts as a consistency test for our
scenario, we observe that $\Om_\nu$ starts growing at the recent
past which is a novel feature that the scenario at hand brings in.

\begin{figure}[h]
\centering
\includegraphics[scale=.7]{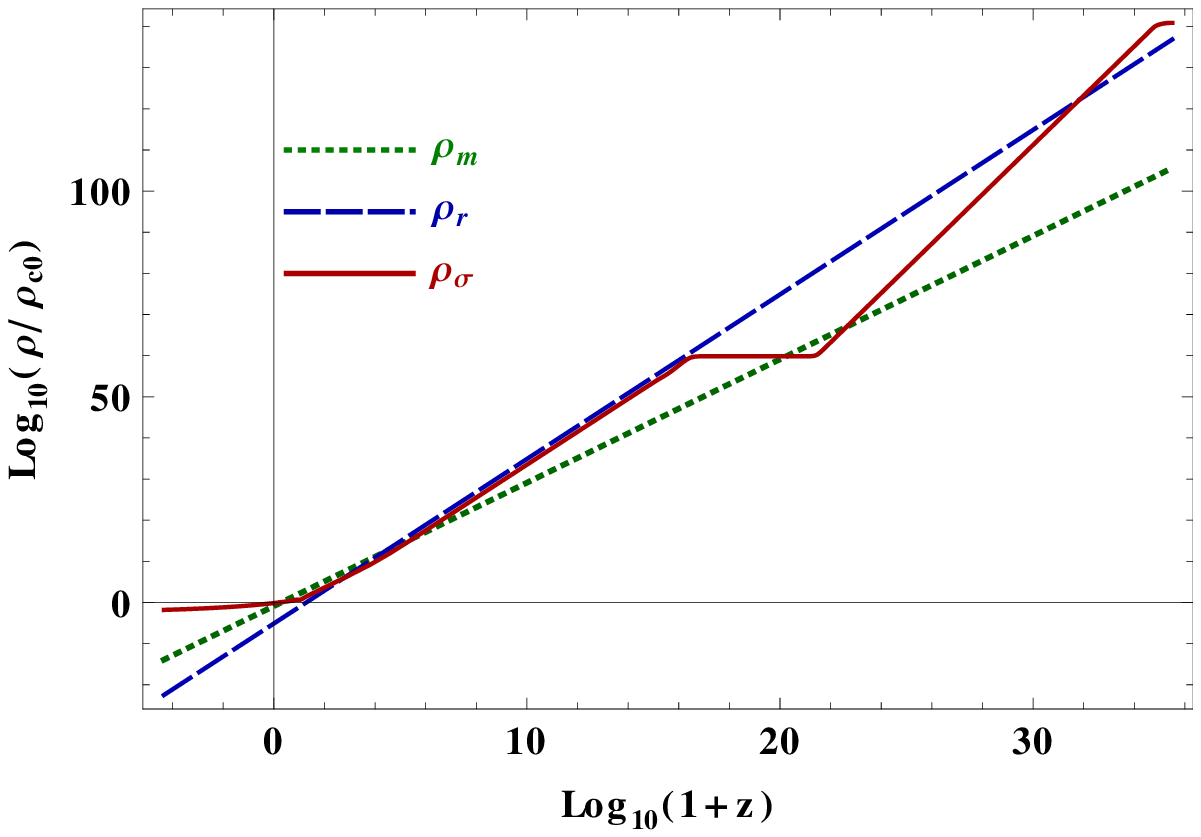}
\includegraphics[scale=.7]{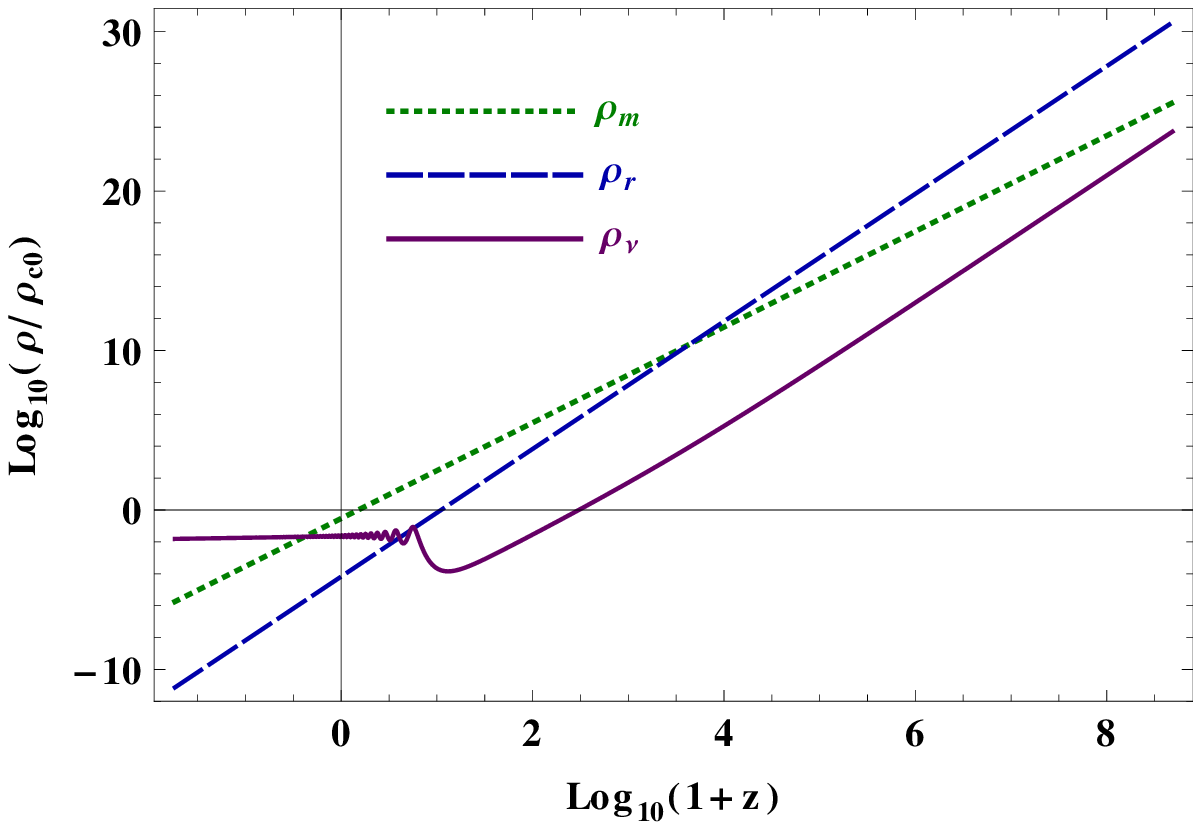}
\caption{Evolutions of different energy densities ($\rho$).
$\rho_r$ (Blue dashed), $\rho_m$ (Green dot-dashed),
$\rho_\sig$ (Red solid (upper panel)), $\rho_\nu$
(Purple solid (lower panel))
represent the densities of radiation, matter,
scalar field $\sigma$ and neutrino. $\rho_{c0}$ is the critical
energy density of universe at present. 
This figures show tracker behavior of the scalar field which trackes
radiation and matter up to recent past and then takes over matter
and becomes dominant component of the universe.
The figure in the lower panel shows that at late times when
neutrinos become non-relativistic, $\rho_\nu$ takes over radiation
and slowly grows thereafter. At the present epoch $\rho_\nu$ is
still sub-dominant but would take over matter in the future.
To plot this figure we have considered $\al=10$, $\tilde{\gamma}=30$ and
$z_{\rm dur}=10$. Since at the end of inflation $k_{\rm end}=\al/\sqrt{2}$, we can take the
initial value of $\lam\sim\mathcal{O}(1)$.
}
\label{fig:rho}
\end{figure}

In Fig.~\ref{fig:rho}, we present the post inflationary evolution of
the energy densities: ($\rho_m$), radiation ($\rho_r$), neutrinos
($\rho_\nu$) and scalar field ($\rho_\sig$). Figure shows that field
energy density soon after the end of inflation enters the kinetic
regime which is attributed to the steep behavior of the potential.
Initially scalar field energy density is much larger than that of
radiation, the field therefore overshoots the background and
freezes. It remains in the locking regime till the radiation density
becomes of the order of field energy density. The field then begins
evolving and tracks radiation and matter. At late times, the field
takes  over matter and becomes dominant component of the universe.
Let us notice the important role played by the neutrino matter.
 Since neutrinos become non-relativistic at the
recent past, the interaction between neutrinos and field becomes non
zero. Because of this interaction term, the field effective
potential acquires a minimum(Eq.~(\ref{eq:pot_eff_phi})) and field
eventually settles in that minimum of the effective potential  which
causes the scalar field to exit from  scaling regime to de Sitter
phase. As the neutrino mass settles to its present value, the
numerical value of $V_{\rm eff}$ in the minimum is of the order of
present value of dark energy provided we choose the parameter
${\t\gam}$ appropriately; no much fine tuning is involved in this
process. Fig.~\ref{fig:rho}, therefore, presents the desired post
evolutionary evolution of our universe.

\begin{figure}[!]
\centering
\includegraphics[scale=.7]{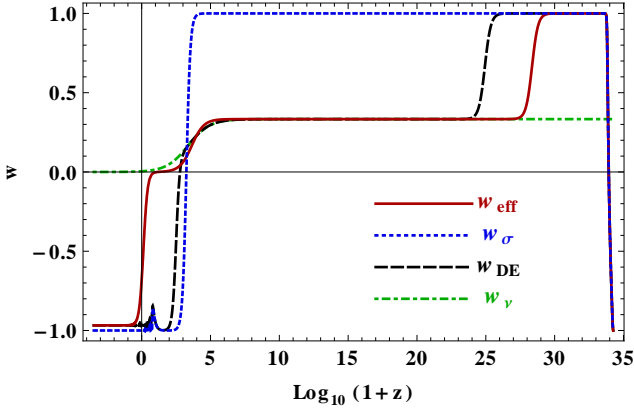}
\caption{Figure shows the evolutions of  equation of state
parameters ($w$). $w_\sig$ (Blue dotted), $w_{\rm DE}$ (Black
dashed), $w_{\rm eff}$ (Red solid) and $w_\nu$ (Green dot-dashed)
represent scalar field $\sigma$, dark energy, effective and neutrino
equation of states respectively and are represented by the Eqs.
(\ref{eq:w_sig}), (\ref{eq:w_DE}), (\ref{eq:w_eff}) and
(\ref{eq:w_nu}). This figure clearly shows that at the present time
$w_\sig$ and $w_{\rm DE}$ are very close to $-1$. To plot this
figure we have considered $\al=10$, $\tilde{\gamma}=30$ and $z_{\rm
dur}=3.6$. Since at the end of inflation $k_{\rm end}=\al/\sqrt{2}$
we can have taken the initial value of $\lam\sim\mathcal{O}(1)$.}
\label{fig:eos}
\end{figure}

In Fig.~\ref{fig:eos} we depict the evolution of the various
equation-of-state parameters. As we observe, during the
radiation dominated era $w_r=1/3$ and $w_\nu=1/3$, and in the recent universe
 $w_\sig, w_{\rm DE} ~{\rm and}~ w_{\rm eff}\sim -1$, that is the dark-energy
component behaves like a cosmological constant.

\begin{figure}[h]
\centering
\includegraphics[scale=.8]{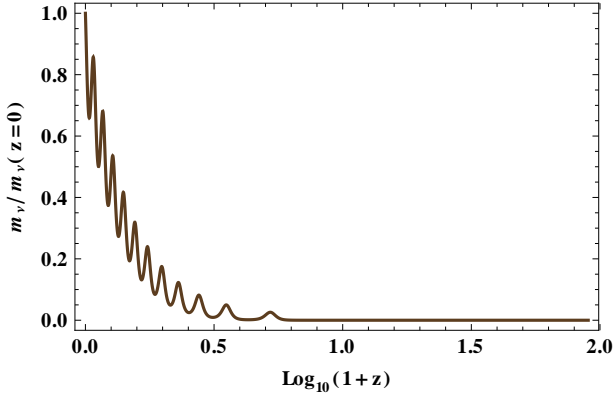}
\caption{Evolution of normalized neutrino mass ($m_\nu/m_\nu(z=0)$)
is shown versus the redshift in this figure. We have used the numerical
values, $\al=10$, $\tilde{\gamma}=30$ and $z_{\rm dur}=3.6$ for
plotting the figure} \label{fig:nu_mass}
\end{figure}

Last but not least, for completeness,  we show in Fig.
\ref{fig:nu_mass}, the evolution of the growing neutrino mass
(normalized with its present value). When neutrinos are relativistic
they behave like radiation and the interaction term between neutrino
and field is zero thereby the mass ratio is constant. In the recent
past ($z\sim 4-10$), neutrinos become non-relativistic and the
interaction term builds up giving rise to the growth of neutrino
mass.

\subsection{Asymptotic behavior: fixed points and stability issues}
\label{Asymptotic behavior}

In order to reveal the late-time behavior of the scenario at hand,
in this subsection, we perform a detailed phase-space analysis of
the cosmological equations (\ref{eq:Fried1}), (\ref{eq:Fried2}),
(\ref{eq:eom_sig}),(\ref{eq:mnu_sig}),
(\ref{DEevol}),(\ref{eq:w_nu}). In this way we can bypass the
complexities of the cosmological equations,  which do not allow for
a compete analytical treatment and extract the late-time, asymptotic
behavior of the universe.

In order to transform the cosmological equations into an autonomous system,
we define the dimensionless auxiliary variables
\begin{eqnarray}
 x &=& \frac{\dot \sig}{\sqrt{6}H\Mpl} \, ,
 \label{eq:x}\\
 y &=& \frac{\sqrt{V}}{\sqrt{3}H\Mpl} \, ,
 \label{eq:y}\\
 \lambda &=& -\frac{\Mpl}{V(\sigma)}\frac{ dV(\sigma)}{ d\sig}
=-\frac{\Mpl}{k(\phi)}\frac{1}{V(\phi)}\frac{\partial
V(\phi)}{\partial\phi} =\frac{\al}{k(\phi)},\ \ \ \
 \label{eq:lam}
\end{eqnarray}
where in the last definition we used relation (\ref{dVsigma}), and
the $k(\phi)$-term is given by (\ref{kphi}). In order to simplify
our analysis, we shall use approximations valid at late times. Since
in this section we are dealing with late-time cosmology ($\chi\gg m$
or equivalently $\phi\gg -2 \Mpl\ln (\mu_m)/\al$), instead of the
full $k(\phi)$, we can use its late-time approximate value.
Expanding (\ref{kphi}) and
 keeping up to first order in $e^{-\al\phi/\Mpl}$, we find
\begin{eqnarray}
 k^2(\phi)\approx 1+\frac{\al^2-\t\al^2}{\t\al^2 \mu_m^2}e^{-\al\phi/\Mpl}
\,,
\end{eqnarray}
which satisfies the discussed requirements that after the end of
inflation $k^2(\phi)$ goes rapidly towards 1 for $\al>\t\al$ and
$\t\al\ll 1$ (note that we could have used  even more approximate
expression (\ref{k=1}), namely $ k^2(\phi)\approx1$ , since for
post-inflationary evolution this approximation is also very close to
the exact behavior that arises from the exact numerical evolution of
the cosmological system). Thus, the auxiliary variable $\lambda$
from (\ref{eq:lam}) becomes
\begin{eqnarray}
\lambda=\alpha\left[ 1+\frac{\al^2-\t\al^2}{\t\al^2
\mu_m^2}e^{-\al\phi/\Mpl}\right]^{-1/2}
\,.
\end{eqnarray}

Additionally, in order to compare with observations, we will use  the
dimensionless density parameters $ \Omega_m
$,$\Omega_r$,$\Omega_\nu$,$ \Omega_\sig$ given by
(\ref{eq:Omega_m}-(\ref{eq:Omega_m}.

In summary, using the six dimensionless variables
$w_\nu$, $x$, $y$, $\lambda$, $\Omega_m$ and $\Omega_r$, we can transform
our cosmological system  of
equations
(\ref{eq:Fried1}),(\ref{eq:Fried2}),(\ref{eq:eom_sig}),(\ref{eq:mnu_sig}),
(\ref{DEevol}),(\ref{eq:w_nu}) into its autonomous form:
\begin{eqnarray}
 \frac{{\rm d}x}{{\rm d}N} &=& \frac{x}{2}\(3w_\nu
\Om_\nu+\Om_r-3y^2-3\)+\frac{3x^3}{2}
 +\sqrt{\frac{3}{2}}y^2\lam  \nn \\ &&
+\sqrt{\frac{3}{2}}(3w_\nu-1)\t\gam\lam\Omega_\nu \, ,
  \label{eq:xp} \\
 \frac{{\rm d}y}{{\rm d}N} &=&
\frac{y}{2}\(3x^2-\sqrt{6}x\lam+3+3w_\nu\Om_\nu+\Om_r\) \nn \\ &&
 -\frac{3 y^3}{2} \, ,
  \label{eq:yp} \\
  \frac{{\rm d}\Om_r}{{\rm d}N} &=& -\Om_r\(1-3x^2+3y^2-3w_\nu\Om_\nu-\Om_r\)
\, ,
   \label{eq:Omega_rp} \\
   \frac{{\rm d}\Om_m}{{\rm d}N} &=& \Om_m\(3x^2-3y^2+3w_\nu\Om_\nu+\Om_r\)
\, ,
    \label{eq:Omega_mp} \\
 \frac{{\rm d}w_\nu}{{\rm d}N} &=& \frac{2w_\nu}{ z_{\rm dur}}\(3w_\nu-1\)
\, ,
 \label{eq:wnup}\\
 \frac{{\rm d}\lam}{{\rm d}N} &=&
\sqrt{\frac{3}{2}}x\lam^2\(1-\frac{\lam^2}{\al^2}\) \, ,
 \label{eq:lamp}
\end{eqnarray}
where $N=\ln a$.

\begin{table*}[ht]
\caption[crit]{Fixed points with their nature of stability and eigenvalues
for the autonomous system
(\ref{eq:xp})-(\ref{eq:lamp}) are given here. We always consider here
$z_{\rm dur}>0$ to get the proper behavior of $w_\nu$. Values of the field
equation of states $w_\sig$,
dark energy equation of states $w_{\rm DE}$ and effective equation of states
$w_{\rm eff}$
corresponding to each fixed points are also listed.
 Here
$A=\sqrt{72-16\al^4\t\gam(1+\t\gam)^2+3\al^2(-7+4\t\gam(3+5\t\gam))}$, and
``arbitr''  stands for ``arbitrary''.   }
\begin{center}
\resizebox{\textwidth}{!}{%
\begin{tabular}{ccccccccccccccc}
\hline \hline Cr.P.& $x$ & $y$ &$\lambda$ & $\Om_r$ & $\Om_m$ &
$\Om_\nu$  &$w_\nu$ &$w_\sig$& $w_{\rm DE}$ &
 $w_{\rm eff}$ & Stability   & Eigenvalues\\
\hline

$P_1$   & $1$& $0$&$\al$ & $0$& $0$& $0$&$\frac{1}{3}$ & $1$& $1$& $1$&
Unstable node for $\al<0$ & $3,2,2$,$\frac{2}{z_{\rm
dur}}$,$-\sqrt{6}\al$,$3-\sqrt{\frac{3}{2}}\al$\\
&&&&&&&&&&&Saddle node for $\al>0$ &\\

$P_2$   & $-1$ &$0$&$\al$ & $0$& $0$&$0$&$\frac{1}{3}$ & $1$& $1$& $1$&
Unstable node for $\al>0$ & $3,2,2$,$\frac{2}{z_{\rm
dur}}$,$\sqrt{6}\al$,$3+\sqrt{\frac{3}{2}}\al$\\
&&&&&&&&&&& Saddle for $\al<0$ &\\

$P_3^{\pm}$ & $\frac{\al}{\sqrt{6}}$ & $\pm\sqrt{1-\frac{\al^2}{6}}$
&$\al$&$0$
& $0$&$0$&$\frac{1}{3}$ &
$\frac{\al^2}{3}-1$ & $\frac{\al^2}{3}-1$ & $\frac{\al^2}{3}-1$ &
Saddle & $\frac{2}{z_{\rm dur}},(\al^2-6)/2,\al^2-4,\al^2-4,$\\
&&&&&&&&&& & & $\al^2-3,-\al^2$\\

$P_4^{\pm}$ & $\frac{2\sqrt{2}}{\sqrt{3}\al}$ &
$\pm\frac{2}{\sqrt{3}\al}$&$\al$ & $0$& $0$& $1-\frac{4}{\al^2}$
&$\frac{1}{3}$ &
$\frac{1}{3}$ & $\frac{1}{3}$ & $\frac{1}{3}$ &
Saddle & $-4,1,0,\frac{2}{z_{\rm dur}},-\frac{\al+\sqrt{64-15\al^2}}{2\al}$\\
&&&&&&&&&&  && $-\frac{\al-\sqrt{64-15\al^2}}{2\al}$\\

$P_5^{\pm}$ & $\frac{\sqrt{3}}{\sqrt{2}\al}$ &
$\pm\frac{\sqrt{3}}{\sqrt{2}\al}$&$\al$ & $0$& $1-\frac{3}{\al^2}$ &
0&$\frac{1}{3}$ &
$0$ & $0$ & $0$ &
Saddle & $-3,-1,-1,\frac{2}{z_{\rm dur}},-\frac{3\left(\al+\sqrt{24-7
\al^2}\right)}{4\al}$\\
&&&&&&&&&&& ~ & $-\frac{3\left(\al-\sqrt{24-7\al^2}\right)}{4\al}$\\

$P_6^{\pm}$ & $\frac{2\sqrt{2}}{\sqrt{3}\al}$ &
$\pm\frac{2}{\sqrt{3}\al}$&$\al$ & $\Om_r$ & $0$& $1-\frac{4}{\al^2}-\Om_r$
&$\frac{1}{3}$   &
$\frac{1}{3}$ & $\frac{1}{3}$ & $\frac{1}{3}$ &
Saddle & $-4,1,0,\frac{2}{z_{\rm dur}},-\frac{\al+\sqrt{64-15\al^2}}{2\al}$\\
&&&&&&&&&&&  & $-\frac{\al-\sqrt{64-15\al^2}}{2\al}$\\

$Q_1$   & $1$& $0$&$\al$ & $0$& $0$& $0$ & $0$& $1$& $1$& $1$&
Saddle & $3,2,-\frac{2}{z_{\rm dur}},-\sqrt{6}\al,3-\sqrt{\frac{3}{2}}\al,$\\
&&&&&&&&&&&& $3+\sqrt{6}\al\t\gam$  \\

$Q_2$   & $-1$ &$0$&$\al$ & $0$& $0$& $0$& $0$ & $1$& $1$& $1$&
Saddle  & $3,2,-\frac{2}{z_{\rm dur}},\sqrt{6}\al,3+\sqrt{\frac{3}{2}}\al,$\\
&&&&&&&&&&&& $3-\sqrt{6}\al\t\gam$ \\

$Q_3^{\pm}$  & $\frac{\al}{\sqrt{6}}$ & $\pm\sqrt{1-\frac{\al^2}{6}}$&$\al$ &
$0$& $0$& $0$& $0$ &
$\frac{\al^2}{3}-1$ & $\frac{\al^2}{3}-1$ & $\frac{\al^2}{3}-1$ &
Stable for $\al^2<\min\{3,\frac{2}{1+\tilde{\gamma}}\}$ & $-\frac{2}{z_{\rm
dur}},-\al^2,\al^2-4,\al^2-3,$\\
&&&&&&&&&&& Saddle otherwise &
$\frac{1}{2}\left(\al^2-6\right),\al^2(1+\t\gam)-3$\\

$Q_4^{\pm}$  & $\frac{2\sqrt{2}}{\sqrt{3}\al}$ &
$\pm\frac{2}{\sqrt{3}\al}$    &$\al$ & $1-\frac{4}{\al^2}$ & $0$& $0$& $0$ &
$\frac{1}{3}$ & $\frac{1}{3}$ & $\frac{1}{3}$ &
Saddle & $-4,1,-\frac{2}{z_{\rm dur}},-\frac{\al+\sqrt{64-15\al^2}}{2\al}$\\
&&&&&&&&&&&  & $-\frac{\al-\sqrt{64-15\al^2}}{2\al},1+4\t\gam$\\

$Q_5^{\pm}$  & $\frac{\sqrt{3}}{\sqrt{2}\al}$ &
$\pm\frac{\sqrt{3}}{\sqrt{2}\al}$  &$\al$ & $0$& $1-\frac{3}{\al^2}$ & $0$&
$0$&
$0$ & $0$ & $0$ &  Stable for $\t\gam<0$     & $-1,3\t\gam,-\frac{2}{z_{\rm
dur}},-\frac{3\left(\al+\sqrt{24-7 \al^2}\right)}{4\al}$\\
&&&&&&&&&&&
 and $\sqrt{3}<\al\leq
2\sqrt{\frac{6}{7}}$ &
$-\frac{3\left(\al-\sqrt{24-7\al^2}\right)}{4\al},-3$\\
&&&&&&&&&&&   or $-2 \sqrt{\frac{6}{7}}\leq
\al<-\sqrt{3}$     & \\
&&&&&&&&&&& Saddle for $\t\gam>0$   & \\

$Q_{6}$   & $-\frac{1}{\sqrt{6}\al\t\gam}$ & $0$ &$\al$ &
$1-\frac{1}{2\al^2\t\gam^2}$ & $0$&
$\frac{1}{3\al^2\t\gam^2}$ & $0$ & $1$& $\frac{1}{3}$ & $\frac{1}{3}$ &
Saddle & $1,-\frac{2}{z_{\rm
dur}},\frac{1}{2}\left(4+\frac{1}{\t\gam}\right),\frac{1}{\t\gam},$\\
&&&&&&&&&&& &
$-\frac{\al\t\gam+\sqrt{2-3\al^2\t\gam^2}}{2\al\t\gam},-\frac{\al\t\gam-\sqrt{
2-3\al^2\t\gam^2}}{2\al\t\gam}$ \\

$Q_7$   & $-\sqrt{\frac{2}{3}}\al\t\gam$ & $0$ &$\al$ & $0$& $0$&
$1-\frac{2\al^2\t\gam^2}{3}$  & $0$&
$1$& $\frac{2\al^2\t\gam^2}{3}$ & $\frac{2\al^2\t\gam^2}{3}$ &
Saddle & $-\frac{2}{z_{\rm
dur}},2\al^2\t\gam,2\al^2\t\gam^2,-\frac{3}{2}+\al^2\t\gam^2,$\\
&&&&&&&&&&&  & $-1+2\al^2\t\gam^2,\frac{3}{2}+\al^2\t\gam(1+\t\gam)$\\

$Q_8^{\pm}$  & $\frac{\sqrt{3}}{\sqrt{2}\al(1+\t\gam)}$ ~ &
$\pm\frac{\sqrt{3+2\al^2\t\gam(1+\t\gam)}}{\sqrt{2}\sqrt{\al^2(1+\t\gam)^2}}$
&$\al$ &
$0$& $0$& $\frac{-3+\al^2(1+\t\gam)}{\al^2(1+\t\gam)^2}$  & $0$&
$-\frac{\al^2\t\gam(1+\t\gam)}{3+\al^2\t\gam(1+\t\gam)}$
& $-\frac{\t\gam}{1+\t\gam}$ & $-\frac{\t\gam}{1+\t\gam}$ &
Attractor for  & $-\frac{2}{z_{\rm
dur}},-\frac{3}{1+\t\gam},-4+\frac{3}{1+\t\gam},-\frac{3\t\gam}{1+\t\gam},$\\
\vspace{0.3cm}
&&&&&&&&&&&  see Fig.~\ref{fig:EV_Q8} &
$\frac{-3\al(1+2\t\gam)+\sqrt{3}A}{4\al(1+\t\gam)},-\frac{3\al(1+2\t\gam)+\sqrt{3}A
}{4\al(1+\t\gam)}$\\

$R_1$   & $0$&$0$& $\lambda$ & $0$& $0$& $1$& $\frac{1}{3}$ & arbitr &
$\frac{1}{3}$ & $\frac{1}{3}$ & Saddle & $2,-1,1,0,0,\frac{2}{z_{\rm dur}}$\\

$R_2$   & $0$&$0$& $\lambda$ & $0$& $1$& $0$& $\frac{1}{3}$ & arbitr & arbitr
&$0$
& Saddle  & $-\frac{3}{2},\frac{3}{2},-1,-1,0,\frac{2}{z_{\rm dur}}$\\

$R_3$   & $0$&$0$& $\lambda$ & $\Om_r$ & $0$& $1-\Om_r$ & $\frac{1}{3}$ &
arbitr
& arbitr & $\frac{1}{3}$ &
Saddle & $2,-1,1,0,0,\frac{2}{z_{\rm dur}}$\\

$R_4$    & $0$&$0$& $\lambda$ & $1$& $0$& $0$& $0$& arbitr & arbitr &
$\frac{1}{3}$
&
Saddle & $2,-1,1,1,0,-\frac{2}{z_{\rm dur}}$\\
\vspace{0.3cm}
$R_5$    & $0$&$0$& $\lambda$ & $0$& $1$& $0$& $0$& arbitr & arbitr & $0$ &
Saddle  &  $-\frac{3}{2},\frac{3}{2},-1,0,0,-\frac{2}{z_{\rm dur}}$\\

$S_{1}$ & $ 1$ & $0$  &$0$& $0$& $0$& $0$& $\frac{1}{3}$ & $1$& $1$&$1$
&
Saddle & $3,3,2,2,0,\frac{2}{z_{\rm dur}}$\\

$S_{2}$ & $- 1$ & $0$  &$0$& $0$& $0$& $0$& $\frac{1}{3}$ & $1$& $1$&$1$
&
Saddle & $3,3,2,2,0,\frac{2}{z_{\rm dur}}$\\

$S_3^{\pm}$ & $0$& $\pm 1$   &$0$& $0$& $0$& $0$& $\frac{1}{3}$ & $-1$ & $-1$
&
$-1$ & Saddle & $-4,-4,-3,-3,0,\frac{2}{z_{\rm dur}}$\\

$S_4$ & $0$& $0$  &$0$ & $0$& $0$& $1$& $\frac{1}{3}$ & arbitr &
$\frac{1}{3}$ &
$\frac{1}{3}$ & Saddle & $2,-1,1,0,0,\frac{2}{z_{\rm dur}}$ \\

$S_{5}$ & $ 1$ & $0$  &$0$ & $0$& $0$& $0$& $0$& $1$& $1$& $1$& Saddle
&
$3,3,3,2,0,-\frac{2}{z_{\rm dur}}$\\

$S_{6}$ & $- 1$ & $0$  &$0$ & $0$& $0$& $0$& $0$& $1$& $1$& $1$& Saddle
&
$3,3,3,2,0,-\frac{2}{z_{\rm dur}}$\\

$S_7^{\pm}$ & $0$& $\pm 1$   &$0$ & $0$& $0$& $0$& $0$& $-1$ & $-1$ & $-1$ &
Stable
& $-4,-3,-3,-3,0,-\frac{2}{z_{\rm dur}}$\\

$S_8$   & $0$& $0$  &$0$ & $0$& $\Om_m$ & $1-\Om_m$ & $0$& arbitr & $0$& $0$&
Saddle
& $-\frac{3}{2},\frac{3}{2},-1,0,0,-\frac{2}{z_{\rm dur}}$ \\

$S_9$   & $0$& $0$  &$0$ & $0$& $0$& $1$& $0$& arbitr & $0$& $0$& Saddle &
$-\frac{3}{2},\frac{3}{2},-1,0,0,-\frac{2}{z_{\rm dur}}$ \\

 \hline\hline
\end{tabular}}
\label{Tablecrit}
\end{center}
\end{table*}

Finally, let us express the remaining observables in terms of the auxiliary
variables $w_\nu$, $x$, $y$, $\lambda$, $\Omega_m$ and $\Omega_r$.
Concerning the density parameters $\Omega_\sig$ and $\Omega_\nu$,
they can be expressed as
\begin{eqnarray}
\Om_\sig=x^2+y^2
 \label{Omsigma}
\end{eqnarray}
and
 \begin{eqnarray}
  \label{Omnu}
 \Om_\nu= 1-\Om_\sig-\Om_m-\Om_r \, ,
 \end{eqnarray}
 where the last expression arises from the Friedmann
equation (\ref{eq:Fried1}). Additionally, according to (\ref{rhoDE}), in the
scenario at hand the effective dark energy density parameter will be just
\begin{eqnarray}
 \Om_{\rm DE}=\Om_\sig+\Om_\nu \, .
 \label{eq:density_DE}
\end{eqnarray}
Lastly, the equation-of-state parameters of the total universe
content, of the scalar-field sector and of the dark-energy sector,
defined in (\ref{eq:w_eff})-(\ref{eq:w_DE}) write as
\begin{eqnarray}
 w_{\rm eff} &=& x^2-y^2+w_\nu\Om_\nu+\frac{\Omega_r}{3} \, ,
 \label{eq:w_eff} \\
 w_{\sig} &=& \frac{x^2-y^2}{x^2+y^2} \, ,
 \label{eq:w_sig} \\
 w_{\rm DE} &=& \frac{w_{\rm eff}-\frac{1}{3}\Om_r}{\Om_{\rm DE}}
 =\frac{x^2-y^2+w_\nu\Om_\nu}{1-\Omega_m-\Omega_r}\, .
 \label{eq:w_DE}
\end{eqnarray}

We first extract the critical points of the above autonomous system by
equating  Eqs.~(\ref{eq:xp})-(\ref{eq:lamp}) to zero. Then in order to
determine their stability properties we follow the usual procedure and we
expand around them, obtaining the perturbation equations in matrix form
\cite{Copeland:1997et,Chen:2008ft,Xu:2012jf,Leon:2012mt}. Thus, the
eigenvalues of of the coefficient-matrix calculated for each critical point,
determine its type and stability.

The real and physically meaningful (that is corresponding to
$0\leq\Omega_{i}\leq1$) critical points  for  $w_\nu$, $x$, $y$,
$\lambda$, $\Omega_m$ and $\Omega_r$  are presented in Table \ref{Tablecrit},
along with their stability conditions and the corresponding eigenvalues of
the perturbation matrix. Additionally, using
(\ref{Omnu}), (\ref{eq:w_eff}),(\ref{eq:w_eff}) and (\ref{eq:w_DE}), for
each critical point we calculate the corresponding values of $\Omega_\nu$,
$ w_{\rm eff}$, $ w_{\sig}$ and $w_{\rm DE}$. Finally, note that points with
$y>0$, that is with $H>0$, correspond to expanding universe, while those
with $y<0$ correspond to a contracting one and we denote them by the index
$-$ in the points name (for $y=0$ the universe can be either contracting or
expanding).

Amongst the critical points, the stable ones are the most
interesting: they are the late time attractors of the dynamics. As
we observe, there are four conditionally stable fixed points (we
focus on the expanding ones):

\begin{itemize}

 \item  Point $Q_3^+$  corresponds to dark-energy dominated
($\Omega_{\rm DE}=\Omega_\sigma+\Omega_\nu=1$), quintessence-like universe
($w_{\rm DE}\geq-1$), which can be accelerating (if $w_{\rm eff}<-1/3$) or not
(if $w_{\rm eff}>-1/3$). As embedded in the model, the neutrinos behave as
dust ($w_\nu=0$). This point is a good candidate for the description of the
late-time universe since it is in agreement with observations.

\item Point $Q_5^+$ corresponds to a universe with $0<\Omega_m<1$ and  $
0<\Omega_{\rm DE}<1$, that is it can alleviate the coincidence problem since
dark-energy and dark-matter density parameters can be of the same order.
However, the fact that it is non-accelerating universe, with a stiff
dark energy equation-of-state parameter, which are not favored by
observations, does not make it a good candidate for the description of the
late-time universe.

\item Point $Q_8^+$ is the novel point of the scenario at hand. It
corresponds to a quintessence-like universe ($w_{\rm DE}\geq-1$), which can be
accelerating (if $w_{\rm eff}<-1/3$, that is if $\t\gam>1/2$) or not.
Additionally, it has $0<\Omega_m<1$ and  $0<\Om_{\rm DE}<1$, that is it can
alleviate the coincidence problem, and the neutrinos behave as dust. The
interesting feature of this point, is that its properties are determined by
the neutrino-depending quantity $\t\gam$, which was not the case in
the other critical points. The region in the $\al-\t\gam$ plane for which
$Q_8^+$is stable is shown in Fig.~\ref{fig:EV_Q8}

 \begin{figure}[h]
\centering
\includegraphics[scale=.7]{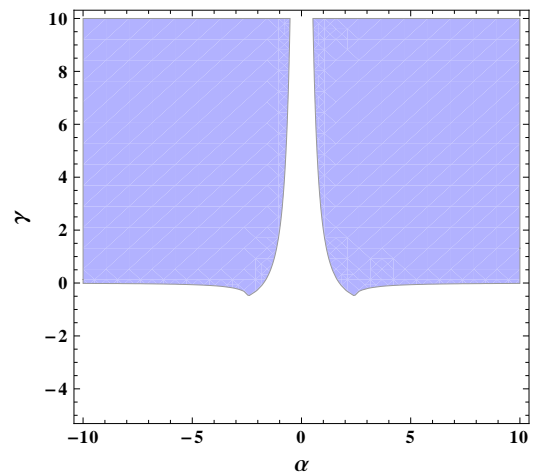}
\caption{Shaded region in the above figure shows the allowed values of
$\al$ and $\t\gam$ for which points $Q_8^\pm$ are stable. The above
regions can be extrapolated for $\al,\t\gam>10$ and $\al,\t\gam<-10$.}
\label{fig:EV_Q8}
\end{figure}

\item Point $S_7^+$ correspond to a de Sitter ($w_{\rm eff}=-1$),
accelerating universe, which is dark-energy dominated  ($\Omega_{\rm DE}=1$),
with the dark energy behaving like a cosmological constant ($w_{\rm DE}=-1$) and
the neutrinos behaving as dust. (Although this point is non-hyperbolic, since
it has one zero eigenvalue amongst the negative ones, an immediate
application o the center manifold \cite{wiggins,Leon:2009rc} analysis shows
that it behaves as stable.)

 \end{itemize}

Finally, note that the points $P_1$ and $P_2$ represent the scalar-field
kinetic-energy dominated regime (the kinetic regime) we mentioned in the
previous subsection.

 Let us make a comment here on the standard-quintessence limit of the
scenario of variable gravity, which acts as a self-consistency test of our
analysis. Clearly, this is obtained when $\lambda=\rm const$, $\Omega_r=0$ and
$\Omega_\nu=0$ (since $\Omega_\nu=0$ the value of $w_\nu$ does not play a
role), that is one freezes these variables to these values, in Table
\ref{Tablecrit}, and thus neglects about the four corresponding eigenvalues
(the system cannot get perturbed in these directions). In this case, we do
recover the standard-quintessence points of
\cite{Copeland:1997et}, and in particular $Q_3^+$ becomes the physically
interesting dark-energy-dominated, quintessence-like point, while  $Q_5^+$
becomes the stiff dark-energy one. However, we mention that in the case
where the standard-quintessence limit is considered, that is one imposes the
above requirements, $P_3^+$ and  $P_5^+$ also coincide with $Q_3^+$
and $Q_5^+$, and thus with the two stable standard-quintessence points.

\begin{figure}[h]
\centering
\includegraphics[scale=.5]{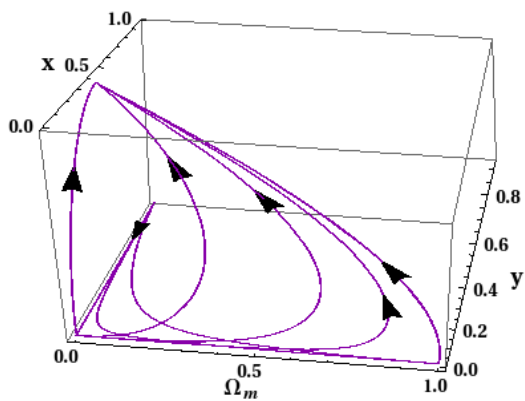}
\caption{Projection of the phase space diagram on $x-y-\Omega_m$ subspace
for the autonomous system
(\ref{eq:xp})-(\ref{eq:lamp}), for $\al=\t\gam=1$ and $\Om_\nu=0$.
 In this case the universe
at late times results to the dark-energy dominated, quintessence-like
universe  $Q_3^+$, which moreover is accelerating for these parameter
values.
}
\label{fig:pp_Q3}
\end{figure}

In order to present the obtained results in a more transparent way, we
perform a numerical elaboration of our cosmological system. In Fig.
\ref{fig:pp_Q3} we depict the
projection of the phase space on $x-y$ plane, for $\al=\t\gam=1$, considering
$\Om_\nu=0$ and $0\leq x^2+y^2\leq 1$. In this case the universe
at late times results to the dark-energy dominated, quintessence-like
universe  $Q_3^+$, which moreover is accelerating for these parameter values.

\begin{figure}[h]
\centering
\includegraphics[scale=.5]{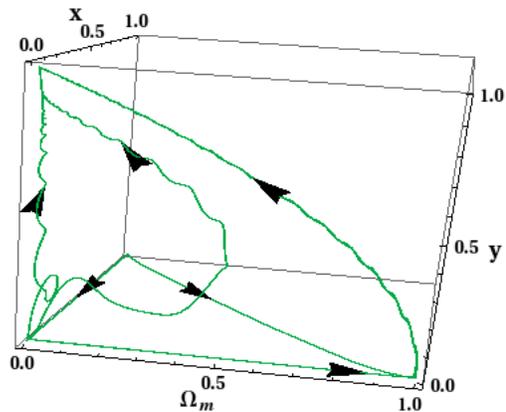}
\caption{Projection of the phase space diagram on $x-y-\Omega_m$ subspace
for the autonomous system
(\ref{eq:xp})-(\ref{eq:lamp}), for $\al=10$ and $\t\gam=30$.
 In this case the universe
at late times is attracted by the quintessence-like, neutrino-depending,
stable point  $Q_8^+$, which moreover is accelerating for these parameter
values.}
\label{fig:pp_Q8}
\end{figure}

Similarly, in Fig.~\ref{fig:pp_Q8} we present the phase-space
evolution   for  $\al=10$ and $\t\gam=30$, in the case  where the universe at
late times is attracted by the
novel stable point $Q_8^+$, that is by a quintessence-like, neutrino-depending
universe, which moreover is accelerating for these parameter values.

\section{Conclusions}
\label{Conclusions}

In the present work we have investigated a scenario of variable
gravity \cite{Wetterich:2013jsa,Wetterich:2013aca} in context with
quintessential inflation$-$
 a unified description of cosmic  evolution from inflation
to radiation, matter and dark-energy epochs. In variable gravity the
Planck mass is driven by a scalar-field, which additionally drives
the mass of the various particles. This field-depending mass,
amongst others, leads to the appearance of an effective interaction
between the scalar field and matter and neutrinos. Furthermore,
through suitable conformal transformations, one can formulate this
model in the Einstein frame  in terms of a canonical scalar field
with an effective non-minimal coupling between the canonical field
and the neutrinos. The cold dark matter is minimally coupled in this
framework. The key assumption in the model is related to the field
dependence of masses in Jordan frame such that cold dark matter and
baryonic matter has standard behavior in Einstein frame whereas the
neutrino masses grow with field in a specific way. The canonical
scalar field at early times is shown to drive inflation with
required number of e-folds $\mathcal{N}$ (which is approximately
equal to $70$ in the model under consideration) and tensor to scalar
ratio of perturbations, $r\simeq 0.11$ consistent with Planck data
within $2\sigma$ confidence level. After inflation, the field
potential fast turns into a steep exponential potential such that
the field enters into kinetic regime with field energy density
redshifting as $a^{-6}$. We checked that gravitational particle
production as a reheating mechanism is
inefficient($(\rho_\phi/\rho_{r})_{\rm end}\simeq 10^{11}$) and it
takes long for the radiative regime to commence. The amplitude of
relic gravity waves enhances during kinetic regime such that
nucleosynthesis constraint($(\rho_\phi/\rho_{r})_{\rm end}\lesssim
10^9$) is violated at the beginning of radiation domination in
this case. We then implemented instant preheating mechanism which
involves the coupling of inflaton with a scalar field $\chi$ such
that $m_\chi=g|\phi|$ which in turn couples to matter field, $h\chi
\bar{\psi}\psi$. At the end of inflation, the mass of $\chi$ changes
non-adiabatically giving rise to $\chi$ production. Assuming that
the energy of $\chi$ production is thermalized,  we can achieve
$(\rho_\phi/\rho_{r})_{\rm end}\lesssim 10^9$ provided that
$g\gtrsim 6\al\times 10^{-5}$. Since after inflation, $\phi$ grows
fast, the produced particles are shown to decay fast into
$\bar{\psi}\psi$ avoiding any back reaction of $\chi$ particles on
the post inflationary dynamics of field $\phi$ provided that
$h\gtrsim 2g^{-1/2}10^{-6}$. We also noticed that particle
production takes place almost instantaneously after the end of
inflation($\phi\leq \phi_p\simeq 4\times 10^{-4}$). We have shown
that instant preheating takes place in a large parameter space (g,h)
and the process is quite efficient to comply with the thermal
history.

Since the field potential in post inflationary era mimics a steep
exponential potential with chosen slope, the field exhibits the
scaling behavior after the locking regime being sub-dominant. At late
times when neutrinos become non-relativistic, the direct coupling of
neutrino matter with scalar field builds up and thanks to
non-minimal coupling, the field potential acquires minimum which
slowly evolves with expansion of universe. The field settles in that
minimum for ever; the transition from scaling regime to late time
cosmic acceleration is successfully triggered by growing neutrino
matter. By performing a detailed phase-space analysis, we showed
that apart from the usual stable attractors similar to those of
standard quintessence, namely the de Sitter, the dark-energy
dominated quintessence-like, and the non-accelerating universe,
stiff-dark ones, the system can result in a new attractor, with
properties depending on the neutrino behavior, corresponding to a
quintessence-like universe.

We have shown that quintessential inflation based upon variable
gravity model can successfully unify inflation and dark energy. The
model based upon instant preheating is shown to be consistent with
observations. The possibility of  detection of the relic gravity
wave background by Advanced LIGO  and LISA are discussed.

As for
early universe, the model complies with the recent Planck data,
showing that within $2\sigma$  confidence level, the scenario is
consistent with observations. The scrutiny of late time
 acceleration and the study of observational constraints on the
 model parameters is deferred to our future work. It
 will also be interesting to carry out detailed investigations of
 stability of neutrino matter under perturbations. One may also
  examine the scenario under consideration in the
 framework of warm inflation which might further improve the tensor
 to scalar ratio of perturbations.


\begin{acknowledgments}
We are grateful to C. Wetterich, S. Mitra and V. Sahni for useful discussions.
MWH acknowledges the local hospitality given by
IUCAA, Pune, India where the part of the work was done.
MWH also acknowledges the funding from CSIR, govt. of India. The research
of ENS is implemented within the framework of the Operational
Program ``Education and Lifelong Learning'' (Actions Beneficiary:
General Secretariat for Research and Technology), and is
co-financed by the European Social Fund (ESF) and the Greek State. MS
thanks the theory division of CERN and ICTP (Italy) where part of the
work was accomplished. He is also thankful to R. Adhikari for useful
discussion on issues related to neutrino matter. 
\end{acknowledgments}


\begin{thebibliography}{150}


\bibitem{Starobinsky:1980te}
  A.~A.~Starobinsky,
  \href{\doi/10.1016/0370-2693(80)90670-X}{Phys.\ Lett.\ B {\bf 91}, 99 (1980)}.

\bibitem{Starobinsky:1982ee}
  A.~A.~Starobinsky,
  \href{\doi/10.1016/0370-2693(82)90541-X}{Phys.\ Lett.\ B {\bf 117}, 175 (1982)}.

\bibitem{Guth:1980zm}
  A.~H.~Guth,
  \href{\doi/10.1103/PhysRevD.23.347}{Phys.\ Rev.\ D {\bf 23}, 347 (1981)}.

\bibitem{Linde:1983gd}
  A.~D.~Linde,
  \href{\doi/10.1016/0370-2693(83)90837-7}{Phys.\ Lett.\ B {\bf 129}, 177 (1983)}.

\bibitem{Linde:1981mu}
  A.~D.~Linde,
  \href{\doi/10.1016/0370-2693(82)91219-9}{Phys.\ Lett.\ B {\bf 108}, 389 (1982)}.

\bibitem{Liddle:1999mq}
  A.~R.~Liddle,
  \href{\arxiv/astro-ph/9901124}{astro-ph/9901124}.

\bibitem{Langlois:2004de}
  D.~Langlois,
  \href{\arxiv/hep-th/0405053}{hep-th/0405053}.

\bibitem{Lyth:1998xn}
  D.~H.~Lyth and A.~Riotto,
  \href{\doi/10.1016/S0370-1573(98)00128-8}{Phys.\ Rept.\  {\bf 314}, 1 (1999)}
  [\href{\arxiv/hep-ph/9807278}{hep-ph/9807278}].

\bibitem{Guth:2000ka}
  A.~H.~Guth,
  \href{\doi/10.1016/S0370-1573(00)00037-5 }{Phys.\ Rept.\  {\bf 333}, 555 (2000)}
  [\href{\arxiv/astro-ph/0002156}{astro-ph/0002156}].

\bibitem{Lidsey:1995np}
  J.~E.~Lidsey, A.~R.~Liddle, E.~W.~Kolb, E.~J.~Copeland, T.~Barreiro and M.~Abney,
  \href{\doi/10.1103/RevModPhys.69.373}{Rev.\ Mod.\ Phys.\  {\bf 69}, 373}
  \href{\doi/10.1103/RevModPhys.69.373}{(1997)}
  [\href{\arxiv/astro-ph/9508078}{astro-ph/9508078}].

\bibitem{Bassett:2005xm}
  B.~A.~Bassett, S.~Tsujikawa and D.~Wands,
  \href{\doi/10.1103/RevModPhys.78.537}{Rev.\ Mod.}\ 
  \href{\doi/10.1103/RevModPhys.78.537}{Phys.\  {\bf 78}, 537 (2006)}
  [\href{\arxiv/astro-ph/0507632}{astro-ph/0507632}].
  
\bibitem{Mazumdar:2010sa} 
  A.~Mazumdar and J.~Rocher,
  \href{\doi/10.1016/j.physrep.2010.08.001}{Phys.\ Rept.\  {\bf 497}, 85}
  \href{\doi/10.1016/j.physrep.2010.08.001}{(2011)}
  [\href{\arxiv/arXiv:1001.0993}{arXiv:1001.0993} [hep-ph]].
  
\bibitem{Wang:2013hva} 
  L.~Wang, E.~Pukartas and A.~Mazumdar,
  \href{\doi/10.1088/1475-7516/2013/07/019}{JCAP {\bf 1307},}
  \href{\doi/10.1088/1475-7516/2013/07/019}{019 (2013)}
  [\href{\arxiv/arXiv:1303.5351}{arXiv:1303.5351} [hep-ph]].
  
\bibitem{Mazumdar:2013gya} 
  A.~Mazumdar and B.~Zaldivar,
  \href{\arxiv/arXiv:1310.5143}{arXiv:1310.5143} [hep-ph].



\bibitem{Copeland:2006wr}
  E.~J.~Copeland, M.~Sami and S.~Tsujikawa,
  \href{\doi/10.1142/S021827180600942X}{Int.\ J.\ Mod.}\ 
  \href{\doi/10.1142/S021827180600942X}{Phys.\ D {\bf 15}, 1753 (2006)}
  [\href{\arxiv/hep-th/0603057}{hep-th/0603057}].

\bibitem{Sahni:1999gb}
  V.~Sahni and A.~A.~Starobinsky,
  \href{\doi/10.1142/S0218271800000542}{Int.\ J.\ Mod.\ Phys.\ D}
  \href{\doi/10.1142/S0218271800000542}{{\bf 9}, 373 (2000)}
  [\href{\arxiv/astro-ph/9904398}{astro-ph/9904398}].

\bibitem{Frieman:2008sn}
  J.~Frieman, M.~Turner and D.~Huterer,
  \href{\doi/10.1146/annurev.astro.46.060407.145243 }{Ann.\ Rev.}\ 
  \href{\doi/10.1146/annurev.astro.46.060407.145243 }{Astron.\ Astrophys.\  {\bf 46}, 385 (2008)}
  [\href{\arxiv/arXiv:0803.0982}{arXiv:0803.0982} [astro-ph]].

\bibitem{Padmanabhan:2002ji}
  T.~Padmanabhan,
  \href{\doi/10.1016/S0370-1573(03)00120-0}{Phys.\ Rept.\  {\bf 380}, 235 (2003)}
  [\href{\arxiv/hep-th/0212290}{hep-th/0212290}].

\bibitem{Padmanabhan:2006ag}
  T.~Padmanabhan,
  \href{\doi/10.1063/1.2399577}{AIP Conf.\ Proc.\  {\bf 861}, 179 (2006)}
  [\href{\arxiv/astro-ph/0603114}{astro-ph/0603114}].

\bibitem{Sahni:2006pa}
  V.~Sahni and A.~Starobinsky,
  \href{\doi/10.1142/S0218271806009704}{Int.\ J.\ Mod.\ Phys.\ D {\bf 15},}
  \href{\doi/10.1142/S0218271806009704}{2105 (2006)}
  [\href{\arxiv/astro-ph/0610026}{astro-ph/0610026}].

\bibitem{Peebles:2002gy}
  P.~J.~E.~Peebles and B.~Ratra,
  \href{\doi/10.1103/RevModPhys.75.559}{Rev.\ Mod.\ Phys.\  {\bf 75},}
  \href{\doi/10.1103/RevModPhys.75.559}{559 (2003)}
  [\href{\arxiv/astro-ph/0207347}{astro-ph/0207347}].

\bibitem{Perivolaropoulos:2006ce}
  L.~Perivolaropoulos,
  \href{\doi/10.1063/1.2348048}{AIP Conf.\ Proc.\  {\bf 848}, 698 (2006)}
  [\href{\arxiv/astro-ph/0601014}{astro-ph/0601014}].

\bibitem{Sami:2009dk}
  M.~Sami,
  \href{\arxiv/arXiv:0901.0756}{arXiv:0901.0756} [hep-th].


\bibitem{Sami:2009jx}
  M.~Sami,
  Curr.\ Sci.\  {\bf 97}, 887 (2009)
  [\href{\arxiv/arXiv:0904.3445}{arXiv:0904.3445} [hep-th]].

\bibitem{Sami:2013ssa}
  M.~Sami and R.~Myrzakulov,
  \href{\arxiv/arXiv:1309.4188}{arXiv:1309.4188} [hep-th].



\bibitem{Peebles:1999fz}
  P.~J.~E.~Peebles and A.~Vilenkin,
  \href{\doi/10.1103/PhysRevD.60.103506}{Phys.\ Rev.\ D}
  \href{\doi/10.1103/PhysRevD.60.103506}{{\bf 60},103506 (1999)}
  [\href{\arxiv/astro-ph/9904396}{astro-ph/9904396}].

\bibitem{Sahni:2001qp}
  V.~Sahni, M.~Sami and T.~Souradeep,
  \href{\doi/10.1103/PhysRevD.65.023518 }{Phys.\ Rev.\ D} 
  \href{\doi/10.1103/PhysRevD.65.023518}{{\bf 65}, 023518 (2001)}
  [\href{\arxiv/gr-qc/0105121}{gr-qc/0105121}].

\bibitem{Sami:2004xk}
  M.~Sami and V.~Sahni,
  \href{\doi/10.1103/PhysRevD.70.083513 }{Phys.\ Rev.\ D {\bf 70}, 083513 (2004)}
  [\href{\arxiv/hep-th/0402086}{hep-th/0402086}].

\bibitem{Copeland:2000hn}
  E.~J.~Copeland, A.~R.~Liddle and J.~E.~Lidsey,
  \href{\doi/10.1103/PhysRevD.64.023509}{Phys.}\ 
  \href{\doi/10.1103/PhysRevD.64.023509}{Rev.\ D {\bf 64}, 023509 (2001)}
  [\href{\arxiv/astro-ph/0006421}{astro-ph/0006421}].

\bibitem{Huey:2001ae}
  G.~Huey and J.~E.~Lidsey,
  \href{\doi/10.1016/S0370-2693(01)00808-5}{Phys.\ Lett.\ B {\bf 514}, 217}
  \href{\doi/10.1016/S0370-2693(01)00808-5}{(2001)}
  [\href{\arxiv/astro-ph/0104006}{astro-ph/0104006}].

\bibitem{Majumdar:2001mm}
  A.~S.~Majumdar,
  \href{\doi/10.1103/PhysRevD.64.083503}{Phys.\ Rev.\ D {\bf 64}, 083503 (2001)}
  [\href{\arxiv/astro-ph/0105518}{astro-ph/0105518}].



\bibitem{Dimopoulos:2000md}
  K.~Dimopoulos,
  \href{\doi/10.1016/S0920-5632(01)01058-1}{Nucl.\ Phys.\ Proc.\ Suppl.\  {\bf 95}, 70 (2001)}
  [\href{\arxiv/astro-ph/0012298}{astro-ph/0012298}].

\bibitem{Sami:2003my}
  M.~Sami, N.~Dadhich and T.~Shiromizu,
  \href{\doi/10.1016/j.physletb.2003.07.001}{Phys.\ Lett.\ B}
  \href{\doi/10.1016/j.physletb.2003.07.001}{{\bf 568}, 118 (2003)}
  [\href{\arxiv/hep-th/0304187}{hep-th/0304187}].

\bibitem{Dimopoulos:2002hm}
  K.~Dimopoulos,
  \href{\doi/10.1103/PhysRevD.68.123506}{Phys.\ Rev.\ D {\bf 68}, 123506 (2003)}
  [\href{\arxiv/astro-ph/0212264}{astro-ph/0212264}].

\bibitem{Dias:2010rg}
  M.~Dias and A.~R.~Liddle,
  \href{\doi/10.1103/PhysRevD.81.083515}{Phys.\ Rev.\ D {\bf 81}, 083515}
  \href{\doi/10.1103/PhysRevD.81.083515}{(2010)}
  [\href{\arxiv/arXiv:1002.3703}{arXiv:1002.3703} [astro-ph.CO]].

\bibitem{BasteroGil:2009eb}
  M.~Bastero-Gil, A.~Berera, B.~M.~Jackson and A.~Taylor,
  \href{\doi/10.1016/j.physletb.2009.06.025}{Phys.\ Lett.\ B {\bf 678}, 157 (2009)}
  [\href{\arxiv/arXiv:0905.2937}{arXiv:0905.2937} [hep-ph]].

\bibitem{Chun:2009yu}
  E.~J.~Chun, S.~Scopel and I.~Zaballa,
  \href{\doi/10.1088/1475-7516/2009/07/022}{JCAP {\bf 0907}, 022}
  \href{\doi/10.1088/1475-7516/2009/07/022}{(2009)}
  [\href{\arxiv/arXiv:0904.0675}{arXiv:0904.0675} [hep-ph]].

\bibitem{Bento:2008yx}
  M.~C.~Bento, R.~G.~Felipe and N.~M.~C.~Santos,
  \href{\doi/10.1103/PhysRevD.77.123512}{Phys.}\ 
  \href{\doi/10.1103/PhysRevD.77.123512}{Rev.\ D {\bf 77}, 123512 (2008)}
  [\href{\arxiv/arXiv:0801.3450}{arXiv:0801.3450} [astro-ph]].

\bibitem{Matsuda:2007ax}
  T.~Matsuda,
  \href{\doi/10.1088/1475-7516/2007/08/003}{JCAP {\bf 0708}, 003 (2007)}
  [\href{\arxiv/arXiv:0707.1948}{arXiv:0707.1948} [hep-ph]].

\bibitem{da Silva:2007vt}
  L.~F.~P.~da Silva and J.~E.~Madriz Aguilar,
  \href{\doi/10.1142/S0217732308025747}{Mod.\ Phys.}\ 
  \href{\doi/10.1142/S0217732308025747}{Lett.\ A {\bf 23}, 1213 (2008)}
  [\href{\arxiv/arXiv:0707.0669}{arXiv:0707.0669} [gr-qc]].

\bibitem{Neupane:2007mu}
  I.~P.~Neupane,
  \href{\doi/10.1088/0264-9381/25/12/125013}{Class.\ Quant.\ Grav.\  {\bf 25}, 125013 (2008)}
  [\href{\arxiv/arXiv:0706.2654}{arXiv:0706.2654} [hep-th]].

\bibitem{Dimopoulos:2007bp}
  K.~Dimopoulos,
  \href{\arxiv/hep-ph/0702018}{hep-ph/0702018} [HEP-PH].

\bibitem{Gardner:2007ib}
  C.~L.~Gardner,
  \href{\arxiv/hep-ph/0701036}{hep-ph/0701036}.

\bibitem{Rosenfeld:2006hs}
  R.~Rosenfeld and J.~A.~Frieman,
  \href{\doi/10.1103/PhysRevD.75.043513}{Phys.\ Rev.\ D {\bf 75},}
  \href{\doi/10.1103/PhysRevD.75.043513}{043513 (2007)}
  [\href{\arxiv/astro-ph/0611241}{astro-ph/0611241}].

\bibitem{Bueno Sanchez:2006ah}
  J.~C.~Bueno Sanchez and K.~Dimopoulos,
  \href{\doi/10.1088/1475-7516/2007/10/002}{JCAP {\bf 0710},}
  \href{\doi/10.1088/1475-7516/2007/10/002}{002 (2007)}
  [\href{\arxiv/hep-th/0606223}{hep-th/0606223}].

\bibitem{Membiela:2006rj}
  A.~Membiela and M.~Bellini,
  \href{\doi/10.1016/j.physletb.2006.08.043}{Phys.\ Lett.\ B {\bf 641}, 125}
  \href{\doi/10.1016/j.physletb.2006.08.043}{(2006)}
  [\href{\arxiv/gr-qc/0606119}{gr-qc/0606119}].

\bibitem{Bueno Sanchez:2006eq}
  J.~C.~Bueno Sanchez and K.~Dimopoulos,
  \href{\doi/10.1016/j.physletb.2006.09.070}{Phys.\ Lett.}\ 
  \href{\doi/10.1016/j.physletb.2006.09.070}{B {\bf 642}, 294 (2006)}
  [\href{\doi/10.1016/j.physletb.2006.09.045}{Erratum-ibid.\ B {\bf 647}, 526 (2007)}]
  [\href{\arxiv/hep-th/0605258}{hep-th/0605258}].

\bibitem{Cardenas:2006py}
  V.~H.~Cardenas,
  \href{\doi/10.1103/PhysRevD.73.103512}{Phys.\ Rev.\ D {\bf 73}, 103512 (2006)}
  [\href{\arxiv/gr-qc/0603013}{gr-qc/0603013}].

\bibitem{Zhai:2005ub}
  X.~-h.~Zhai and Y.~-b.~Zhao,
  \href{\doi/10.1088/1009-1963/15/10/046}{Chin.\ Phys.\  {\bf 15}, 2465}
  \href{\doi/10.1088/1009-1963/15/10/046}{(2006)}
  [\href{\arxiv/astro-ph/0511512}{astro-ph/0511512}].

\bibitem{Rosenfeld:2005mt}
  R.~Rosenfeld and J.~A.~Frieman,
  \href{\doi/10.1088/1475-7516/2005/09/003}{JCAP {\bf 0509}, 003}
  \href{\doi/10.1088/1475-7516/2005/09/003}{(2005)}
  [\href{\arxiv/astro-ph/0504191}{astro-ph/0504191}].

\bibitem{Giovannini:2003jw}
  M.~Giovannini,
  \href{\doi/10.1103/PhysRevD.67.123512}{Phys.\ Rev.\ D {\bf 67}, 123512 (2003)}
  [\href{\arxiv/hep-ph/0301264}{hep-ph/0301264}].

\bibitem{Dimopoulos:2002ug}
  K.~Dimopoulos,
  \href{\arxiv/astro-ph/0210374}{astro-ph/0210374}.

\bibitem{Nunes:2002wz}
  N.~J.~Nunes and E.~J.~Copeland,
  \href{\doi/10.1103/PhysRevD.66.043524}{Phys.\ Rev.\ D {\bf 66},}
  \href{\doi/10.1103/PhysRevD.66.043524}{043524 (2002)}
  [\href{\arxiv/astro-ph/0204115}{astro-ph/0204115}].

\bibitem{Dimopoulos:2001qu}
  K.~Dimopoulos,
  \href{\arxiv/astro-ph/0111500}{astro-ph/0111500}.

\bibitem{Dimopoulos:2001ix}
  K.~Dimopoulos and J.~W.~F.~Valle,
  \href{\doi/10.1016/S0927-6505(02)00115-9}{Astropart.\ Phys.}\ 
  \href{\doi/10.1016/S0927-6505(02)00115-9}{{\bf 18}, 287 (2002)}
  [\href{\arxiv/astro-ph/0111417}{astro-ph/0111417}].

\bibitem{Yahiro:2001uh}
  M.~Yahiro, G.~J.~Mathews, K.~Ichiki, T.~Kajino and M.~Orito,
  \href{\doi/10.1103/PhysRevD.65.063502}{Phys.\ Rev.\ D {\bf 65}, 063502 (2002)}
  [\href{\arxiv/astro-ph/0106349}{astro-ph/0106349}].

\bibitem{Kaganovich:2000fc}
  A.~B.~Kaganovich,
  \href{\doi/10.1103/PhysRevD.63.025022}{Phys.\ Rev.\ D {\bf 63}, 025022 (2000)}
  [\href{\arxiv/hep-th/0007144}{hep-th/0007144}].

\bibitem{Peloso:1999dm}
  M.~Peloso and F.~Rosati,
  \href{\doi/10.1088/1126-6708/1999/12/026}{JHEP {\bf 9912}, 026 (1999)}
  [\href{\arxiv/hep-ph/9908271}{hep-ph/9908271}].

\bibitem{Baccigalupi:1998mn}
  C.~Baccigalupi and F.~Perrotta,
  \href{\arxiv/astro-ph/9811385}{astro-ph/9811385}.

\bibitem{Hinterbichler:2013we} 
  K.~Hinterbichler, J.~Khoury, H.~Nastase and R.~Rosenfeld,
  \href{\doi/}{JHEP {\bf 1308}, 053 (2013)}
  [\href{\arxiv/arXiv:1301.6756}{arXiv:1301.6756} [hep-th]].




\bibitem{Kofman:1994rk}
  L.~Kofman, A.~D.~Linde and A.~A.~Starobinsky,
  \href{\doi10.1103/PhysRevLett.73.3195}{Phys.}\ 
  \href{\doi10.1103/PhysRevLett.73.3195}{Rev.\ Lett.\  {\bf 73}, 3195 (1994)}
  [\href{\arxiv/hep-th/9405187}{hep-th/9405187}].

\bibitem{Kofman:1997yn}
  L.~Kofman, A.~D.~Linde and A.~A.~Starobinsky,
  \href{\doi/10.1103/PhysRevD.56.3258}{Phys.\ Rev.\ D {\bf 56}, 3258 (1997)}
  [\href{\arxiv/hep-ph/9704452}{hep-ph/9704452}].

\bibitem{Dolgov:1982th}
  A.~D.~Dolgov and A.~D.~Linde,
  \href{\doi/10.1016/0370-2693(82)90292-1}{Phys.\ Lett.\ B {\bf 116}, 329}
  \href{\doi/10.1016/0370-2693(82)90292-1}{(1982)}.

\bibitem{Abbott:1982hn}
  L.~F.~Abbott, E.~Farhi and M.~B.~Wise,
  \href{\doi/10.1016/0370-2693(82)90867-X}{Phys.\ Lett.\ B}
  \href{\doi/10.1016/0370-2693(82)90867-X}{{\bf 117}, 29 (1982)}.

\bibitem{Ford:1986sy}
  L.~H.~Ford,
  \href{\doi/10.1103/PhysRevD.35.2955}{Phys.\ Rev.\ D {\bf 35}, 2955 (1987)}.

\bibitem{Spokoiny:1993kt}
  B.~Spokoiny,
  \href{\doi/10.1016/0370-2693(93)90155-B}{Phys.\ Lett.\ B {\bf 315}, 40 (1993)}
  [\href{\arxiv/gr-qc/9306008}{gr-qc/9306008}].

\bibitem{Shtanov:1994ce}
  Y.~Shtanov, J.~H.~Traschen and R.~H.~Brandenberger,
  \href{\doi/10.1103/PhysRevD.51.5438}{Phys.\ Rev.\ D {\bf 51}, 5438 (1995)}
  [\href{\arxiv/hep-ph/9407247}{hep-ph/9407247}].

\bibitem{Campos:2002yk}
  A.~H.~Campos, H.~C.~Reis and R.~Rosenfeld,
  \href{\doi/10.1016/j.physletb.2003.09.064}{Phys.}\ 
  \href{\doi/10.1016/j.physletb.2003.09.064}{Lett.\ B {\bf 575}, 151 (2003)}
  [\href{\arxiv/hep-ph/0210152}{hep-ph/0210152}].




\bibitem{Grishchuk:1974ny}
  L.~P.~Grishchuk,
  Sov.\ Phys.\ JETP {\bf 40}, 409 (1975)
  [Zh.\ Eksp.\ Teor.\ Fiz.\  {\bf 67}, 825 (1974)].

\bibitem{Grishchuk:1977zz}
  L.~P.~Grishchuk,
  \href{\doi/10.1111/j.1749-6632.1977.tb37064.x}{Annals N.\ Y.\ Acad.\ Sci.\  {\bf 302}, 439}
  \href{\doi/10.1111/j.1749-6632.1977.tb37064.x}{(1977)}.

\bibitem{Starobinsky:1979ty}
  A.~A.~Starobinsky,
  \href{http://www.jetpletters.ac.ru/ps/1370/article_20738.shtml}{JETP Lett.\  {\bf 30}, 682 (1979)}
  [Pisma Zh.\ Eksp.\ Teor.\ Fiz.\  {\bf 30}, 719 (1979)].

\bibitem{Sahni:1990tx}
  V.~Sahni,
  \href{\doi/10.1103/PhysRevD.42.453}{Phys.\ Rev.\ D {\bf 42}, 453 (1990)}.

\bibitem{Souradeep:1992sm}
  T.~Souradeep and V.~Sahni,
  \href{\doi/10.1142/S0217732392002950}{Mod.\ Phys.\ Lett.\ A {\bf 7},}
  \href{\doi/10.1142/S0217732392002950}{3541 (1992)}
  [\href{\arxiv/hep-ph/9208217}{hep-ph/9208217}].

\bibitem{Giovannini:1998bp}
  M.~Giovannini,
  \href{\doi/10.1103/PhysRevD.58.083504}{Phys.\ Rev.\ D {\bf 58}, 083504 (1998)}
  [\href{\arxiv/hep-ph/9806329}{hep-ph/9806329}].

\bibitem{Giovannini:1999bh}
  M.~Giovannini,
  \href{\doi/10.1103/PhysRevD.60.123511}{Phys.\ Rev.\ D {\bf 60}, 123511 (1999)}
  [\href{\arxiv/astro-ph/9903004}{astro-ph/9903004}].

\bibitem{Langlois:2000ns}
  D.~Langlois, R.~Maartens and D.~Wands,
  \href{\doi/10.1016/S0370-2693(00)00957-6}{Phys.\ Lett.}\ 
  \href{\doi/10.1016/S0370-2693(00)00957-6}{B {\bf 489}, 259 (2000)}
  [\href{\arxiv/hep-th/0006007}{hep-th/0006007}].

\bibitem{Kobayashi:2003cn}
  T.~Kobayashi, H.~Kudoh and T.~Tanaka,
  \href{\doi/10.1103/PhysRevD.68.044025}{Phys.\ Rev.}\ 
  \href{\doi/10.1103/PhysRevD.68.044025}{D {\bf 68}, 044025 (2003)}
  [\href{\arxiv/gr-qc/0305006}{gr-qc/0305006}].

\bibitem{Hiramatsu:2003iz}
  T.~Hiramatsu, K.~Koyama and A.~Taruya,
  \href{\doi/10.1016/j.physletb.2003.10.111}{Phys.\ Lett.}\ 
  \href{\doi/10.1016/j.physletb.2003.10.111}{B {\bf 578}, 269 (2004)}
  [\href{\arxiv/hep-th/0308072}{hep-th/0308072}].

\bibitem{Easther:2003re}
  R.~Easther, D.~Langlois, R.~Maartens and D.~Wands,
  \href{\doi/10.1088/1475-7516/2003/10/014}{JCAP {\bf 0310}, 014 (2003)}
  [\href{\arxiv/hep-th/0308078}{hep-th/0308078}].

\bibitem{Brustein:1995ah}
  R.~Brustein, M.~Gasperini, M.~Giovannini and G.~Veneziano,
  \href{\doi/10.1016/0370-2693(95)01128-D}{Phys.\ Lett.\ B {\bf 361}, 45 (1995)}
  [\href{\arxiv/hep-th/9507017}{hep-th/9507017}].

\bibitem{Gasperini:1992dp}
  M.~Gasperini and M.~Giovannini,
  \href{\doi/10.1103/PhysRevD.47.1519}{Phys.\ Rev.\ D {\bf 47},}
  \href{\doi/10.1103/PhysRevD.47.1519}{1519 (1993)}
  [\href{\arxiv/gr-qc/9211021}{gr-qc/9211021}].

\bibitem{Giovannini:1999qj}
  M.~Giovannini,
  \href{\doi/10.1088/0264-9381/16/9/308}{Class.\ Quant.\ Grav.\  {\bf 16}, 2905 (1999)}
  [\href{\arxiv/hep-ph/9903263}{hep-ph/9903263}].

\bibitem{Giovannini:1997km}
  M.~Giovannini,
  \href{\doi/10.1103/PhysRevD.56.3198}{Phys.\ Rev.\ D {\bf 56}, 3198 (1997)}
  [\href{\arxiv/hep-th/9706201}{hep-th/9706201}].

\bibitem{Gasperini:1992pa}
  M.~Gasperini and M.~Giovannini,
  \href{\doi/10.1016/0370-2693(92)90476-K}{Phys.\ Lett.\ B {\bf 282},}
  \href{\doi/10.1016/0370-2693(92)90476-K}{36 (1992)}.

\bibitem{Giovannini:2009kg}
  M.~Giovannini,
  \href{\doi/0.1186/1754-0410-4-1}{PMC Phys.\ A {\bf 4}, 1 (2010)}
  [\href{\arxiv/arXiv:0901.3026}{arXiv:0901.3026} [astro-ph.CO]].

\bibitem{Giovannini:2008zg}
  M.~Giovannini,
  \href{\doi/10.1016/j.physletb.2008.07.107}{Phys.\ Lett.\ B {\bf 668}, 44 (2008)}
  [\href{\arxiv/arXiv:0807.1914}{arXiv:0807.1914} [astro-ph]].

\bibitem{Giovannini:2010yy}
  M.~Giovannini,
  \href{\doi/0.1103/PhysRevD.81.123003}{Phys.\ Rev.\ D {\bf 81}, 123003 (2010)}
  [\href{\arxiv/arXiv:1001.4172}{arXiv:1001.4172} [astro-ph.CO]].

\bibitem{Tashiro:2003qp}
  H.~Tashiro, T.~Chiba and M.~Sasaki,
  \href{\doi/10.1088/0264-9381/21/7/004}{Class.\ Quant.}\ 
  \href{\doi/10.1088/0264-9381/21/7/004}{Grav.\  {\bf 21}, 1761 (2004)}
  [\href{\arxiv/gr-qc/0307068}{gr-qc/0307068}].



\bibitem{Felder:1998vq}
  G.~N.~Felder, L.~Kofman and A.~D.~Linde,
  \href{\doi/10.1103/PhysRevD.59.123523}{Phys.\ Rev.}\ 
  \href{\doi/10.1103/PhysRevD.59.123523}{D {\bf 59}, 123523 (1999)}
  [\href{\arxiv/hep-ph/9812289}{hep-ph/9812289}].

\bibitem{Felder:1999pv}
  G.~N.~Felder, L.~Kofman and A.~D.~Linde,
  \href{\doi/10.1103/PhysRevD.60.103505}{Phys.\ Rev.}\ 
  \href{\doi/10.1103/PhysRevD.60.103505}{D {\bf 60}, 103505 (1999)}
  [\href{\arxiv/hep-ph/9903350}{hep-ph/9903350}].

\bibitem{Campos:2004nc}
  A.~H.~Campos, J.~M.~F.~Maia and R.~Rosenfeld,
  \href{\doi/10.1103/PhysRevD.70.023003}{Phys.}\ 
  \href{\doi/10.1103/PhysRevD.70.023003}{Rev.\ D {\bf 70}, 023003 (2004)}
  [\href{\arxiv/astro-ph/0402413}{astro-ph/0402413}].





\bibitem{Randall:1999vf}
  L.~Randall and R.~Sundrum,
  \href{\doi/10.1103/PhysRevLett.83.4690}{Phys.\ Rev.\ Lett.\  {\bf 83}, 4690}
  \href{\doi/10.1103/PhysRevLett.83.4690}{(1999)}
  [\href{\arxiv/hep-th/9906064}{hep-th/9906064}].

\bibitem{Randall:1999ee}
  L.~Randall and R.~Sundrum,
  \href{\doi/10.1103/PhysRevLett.83.3370}{Phys.\ Rev.\ Lett.\  {\bf 83}, 3370}
  \href{\doi/10.1103/PhysRevLett.83.3370}{(1999)}
  [\href{\arxiv/hep-ph/9905221}{hep-ph/9905221}].




\bibitem{Wetterich:2013jsa}
  C.~Wetterich,
  \href{\doi/10.1103/PhysRevD.89.024005}{Phys.\ Rev.\ D {\bf 89}, 024005 (2014)}
  [\href{\arxiv/arXiv:1308.1019}{arXiv:1308.1019} [astro-ph.CO]].

\bibitem{Wetterich:2013aca}
  C.~Wetterich,
  \href{\doi/10.1016/j.dark.2013.10.002}{Physics of the Dark Universe (2013)},
  [\href{\arxiv/1303.6878}{arXiv:1303.6878} [astro-ph.CO]].

\bibitem{Wetterich:2013wza}
  C.~Wetterich,
  \href{\doi/10.1016/j.physletb.2013.08.023}{Phys.\ Lett.\ B {\bf 726}, 15 (2013)}
  [\href{\arxiv/arXiv:1303.4700}{arXiv:1303.4700} [astro-ph.CO]].
  
\bibitem{Wetterich:2014eaa} 
  C.~Wetterich,
  \href{\arxiv/arXiv:1401.5313}{arXiv:1401.5313} [astro-ph.CO].
  
\bibitem{Wetterich:2014bma} 
  C.~Wetterich,
  \href{\arxiv/arXiv:1402.5031}{arXiv:1402.5031} [astro-ph.CO].
  

\bibitem{Amendola:1999er}
  L.~Amendola,
 \href{\doi/10.1103/PhysRevD.62.043511}{Phys.\ Rev.\ D {\bf 62}, 043511 (2000)}
 [\href{\arxiv/astro-ph/9908023}{astro-ph/9908023}].



\bibitem{Amendola:2007yx}
  L.~Amendola, M.~Baldi and C.~Wetterich,
  \href{\doi/10.1103/PhysRevD.78.023015}{Phys.\ Rev.}\ 
  \href{\doi/10.1103/PhysRevD.78.023015}{D {\bf 78}, 023015 (2008)}
  [\href{\arxiv/arXiv:0706.3064}{arXiv:0706.3064} [astro-ph]].

\bibitem{Wetterich:2007kr}
  C.~Wetterich,
  \href{\doi/10.1016/j.physletb.2007.08.060}{Phys.\ Lett.\ B {\bf 655}, 201 (2007)}
  [\href{\arxiv/arXiv:0706.4427}{arXiv:0706.4427} [hep-ph]].

\bibitem{Pettorino:2010bv}
  V.~Pettorino, N.~Wintergerst, L.~Amendola and C.~Wetterich,
  \href{\doi/}{Phys.\ Rev.\ D {\bf 82}, 123001 (2010)}
  [\href{\arxiv/arXiv:1009.2461}{arXiv:1009.2461}[astro-ph.CO]].

\bibitem{Fardon:2003eh}
  R.~Fardon, A.~E.~Nelson and N.~Weiner,
  \href{\doi/10.1088/1475-7516/2004/10/005}{JCAP {\bf 0410},}
  \href{\doi/10.1088/1475-7516/2004/10/005}{005 (2004)}
  [\href{\arxiv/astro-ph/0309800}{astro-ph/0309800}].

\bibitem{Bi:2003yr}
  X.~-J.~Bi, P.~-h.~Gu, X.~-l.~Wang and X.~-m.~Zhang,
  \href{\doi/10.1103/PhysRevD.69.113007}{Phys.\ Rev.\ D}
  \href{\doi/10.1103/PhysRevD.69.113007}{{\bf 69}, 113007 (2004)}
  [\href{\arxiv/hep-ph/0311022}{hep-ph/0311022}].

\bibitem{Hung:2003jb}
  P.~Q.~Hung and H.~Pas,
  \href{\doi/10.1142/S0217732305016981}{Mod.\ Phys.\ Lett.\ A {\bf 20}, 1209}
  \href{\doi/10.1142/S0217732305016981}{(2005)}
  [\href{\arxiv/astro-ph/0311131}{astro-ph/0311131}].

\bibitem{Peccei:2004sz}
  R.~D.~Peccei,
  \href{\doi/10.1103/PhysRevD.71.023527}{Phys.\ Rev.\ D {\bf 71}, 023527 (2005)}
  [\href{\arxiv/hep-ph/0411137}{hep-ph/0411137}].

\bibitem{Bi:2004ns}
  X.~-J.~Bi, B.~Feng, H.~Li and X.~-m.~Zhang,
  \href{\doi/10.1103/PhysRevD.72.123523}{Phys.\ Rev.}\ 
  \href{\doi/10.1103/PhysRevD.72.123523}{D {\bf 72}, 123523 (2005)}
  [\href{\arxiv/hep-ph/0412002}{hep-ph/0412002}].

\bibitem{Brookfield:2005td}
  A.~W.~Brookfield, C.~van de Bruck, D.~F.~Mota and D.~Tocchini-Valentini,
  \href{\doi/10.1103/PhysRevLett.96.061301}{Phys.\ Rev.\ Lett.\  {\bf 96}, 061301}
  \href{\doi/10.1103/PhysRevLett.96.061301}{(2006)}
  [\href{\arxiv/astro-ph/0503349}{astro-ph/0503349}].

\bibitem{Brookfield:2005bz}
  A.~W.~Brookfield, C.~van de Bruck, D.~F.~Mota and D.~Tocchini-Valentini,
  \href{\doi/10.1103/PhysRevD.73.083515}{Phys.\ Rev.\ D {\bf 73},}
  \href{\doi/10.1103/PhysRevD.73.083515}{083515 (2006)}
  [\href{\doi/10.1103/PhysRevD.76.049901}{Erratum-ibid.\ D {\bf 76}, 049901 (2007)}]
  [\href{\arxiv/astro-ph/0512367}{astro-ph/0512367}].

\bibitem{Bjaelde:2007ki}
  O.~E.~Bjaelde, A.~W.~Brookfield, C.~van de Bruck, S.~Hannestad, D.~F.~Mota, L.~Schrempp and D.~Tocchini-Valentini,
  \href{\doi/10.1088/1475-7516/2008/01/026}{JCAP {\bf 0801}, 026 (2008)}
  [\href{\arxiv/arXiv:0705.2018}{arXiv:0705.2018} [astro-ph]].

\bibitem{Afshordi:2005ym}
  N.~Afshordi, M.~Zaldarriaga and K.~Kohri,
  \href{\doi/10.1103/PhysRevD.72.065024}{Phys.\ Rev.}\ 
  \href{\doi/10.1103/PhysRevD.72.065024}{D {\bf 72}, 065024 (2005)}
  [\href{\arxiv/astro-ph/0506663}{astro-ph/0506663}].


\bibitem{Mota:2008nj}
  D.~F.~Mota, V.~Pettorino, G.~Robbers and C.~Wetterich,
  \href{\doi/10.1016/j.physletb.2008.03.060}{Phys.\ Lett.\ B {\bf 663}, 160 (2008)}
  [\href{\arxiv/arXiv:0802.1515}{arXiv:0802.1515} [astro-ph]].

\bibitem{LaVacca:2012ir}
  G.~La Vacca and D.~F.~Mota,
  \href{\doi/10.1051/0004-6361/201220971}{Astron.\ Astrophys.\  {\bf 560},}
  \href{\doi/10.1051/0004-6361/201220971}{A53 (2013)}
  [\href{\arxiv/arXiv:1205.6059}{arXiv:1205.6059} [astro-ph.CO]].

\bibitem{Collodel:2012bp}
  L.~G.~Collodel and G.~M.~Kremer,
  \href{\doi/10.1134/S0202289312030036}{Grav.\ Cosmol.\  {\bf 18},}
  \href{\doi/10.1134/S0202289312030036}{196 (2012)}
  [\href{\arxiv/arXiv:1203.3061}{arXiv:1203.3061 [gr-qc]}].


\bibitem{Chiba:1999wt}
  T.~Chiba,
 \href{\doi/10.1103/PhysRevD.60.083508}{Phys.\ Rev.\ D {\bf 60}, 083508 (1999)}
 [\href{\arxiv/gr-qc/9903094}{gr-qc/9903094}].

\bibitem{Uzan:1999ch}
  J.~-P.~Uzan,
 \href{\doi/10.1103/PhysRevD.59.123510}{Phys.\ Rev.\ D {\bf 59}, 123510 (1999)}
 [\href{\arxiv/gr-qc/9903004}{gr-qc/9903004}].

\bibitem{Baccigalupi:2000je}
  C.~Baccigalupi, S.~Matarrese and F.~Perrotta,
 \href{\doi/10.1103/PhysRevD.62.123510}{Phys.}\ 
 \href{\doi/10.1103/PhysRevD.62.123510}{Rev.\ D {\bf 62}, 123510 (2000)}
 [\href{\arxiv/astro-ph/0005543}{astro-ph/0005543}].





\bibitem{Bunn:1996py}
  E.~F.~Bunn, A.~R.~Liddle and M.~J.~White, 1,
  \href{\doi/10.1103/PhysRevD.54.R5917}{Phys.}\ 
  \href{\doi/10.1103/PhysRevD.54.R5917}{Rev.\ D {\bf 54}, 5917 (1996)}
  [\href{\arxiv/astro-ph/9607038}{astro-ph/9607038}].

\bibitem{Bunn:1996da}
  E.~F.~Bunn and M.~J.~White, 1,
  \href{\doi/10.1086/303955}{Astrophys.\ J.\  {\bf 480}, 6}
  \href{\doi/10.1086/303955}{(1997)}
  [\href{\arxiv/astro-ph/9607060}{astro-ph/9607060}].


\bibitem{Ade:2013uln}
  P.~A.~R.~Ade {\it et al.}  [Planck Collaboration],
  [\href{\arxiv/arXiv:1303.5082}{arXiv:1303.5082} [astro-ph.CO]].




\bibitem{LIGO}
 LIGO Home Page,
  \url{http://www.ligo.caltech.edu/}.

  \bibitem{aLIGO}
  \url{https://dcc.ligo.org/LIGO-T0900288/public}

  \bibitem{LISA}
  LISA Home Page,
  \url{http://lisa.nasa.gov/}.

    \bibitem{LISA1}
  \url{http://www.srl.caltech.edu/~shane/sensitivity/}.


\bibitem{Copeland:1997et}
  E.~J.~Copeland, A.~R.~Liddle and D.~Wands,
     \href{http://doi.org/10.1103/PhysRevD.57.4686}{Phys.}\ 
     \href{http://doi.org/10.1103/PhysRevD.57.4686}{Rev.\  D {\bf57}, 4686 (1998)}
  [\href{http://arxiv.org/gr-qc/9711068}{gr-qc/9711068}].




\bibitem{Chen:2008ft}
  X.~-m.~Chen, Y.~-g.~Gong and E.~N.~Saridakis,
 \href{\doi/10.1088/1475-7516/2009/04/001}{JCAP}
 \href{\doi/10.1088/1475-7516/2009/04/001}{{\bf 0904}, 001 (2009)},
 [\href{\arxiv/arXiv:0812.1117}{arXiv:0812.1117} [gr-qc]].


\bibitem{Xu:2012jf}
  C.~Xu, E.~N.~Saridakis and G.~Leon,
    \href{http://doi.org/10.1088/1475-7516/2012/07/005}{JCAP {\bf 1207}, 005}
    \href{http://doi.org/10.1088/1475-7516/2012/07/005}{(2012)}
  [\href{\arxiv/arXiv:1202.3781}{arXiv:1202.3781} [gr-qc]].

\bibitem{Leon:2012mt}
  G.~Leon and E.~N.~Saridakis,
      \href{http://doi.org/10.1088/1475-7516/2013/03/025}{JCAP {\bf 1303},
025 (2013)}
  [\href{\arxiv/arXiv:1211.3088}{arXiv:1211.3088} [astro-ph.CO]].



\bibitem{wiggins} S. Wiggins,
 {\it{ Introduction to Applied Nonlinear Dynamical Systems and Chaos}},
Springer, New York (2003).



\bibitem{Leon:2009rc}
  G.~Leon and E.~N.~Saridakis,
        \href{http://doi.org/10.1088/1475-7516/2009/11/006}{JCAP {\bf 0911},
006 (2009)}
  [\href{\arxiv/arXiv:0909.3571}{arXiv:0909.3571}].



















\end{thebibliography}
\end{document}